\documentclass[submission,PhysLectNotes]{SciPost}
\hypersetup{
    colorlinks,
    linkcolor={red!50!black},
    citecolor={blue!50!black},
    urlcolor={blue!80!black}
}

\usepackage[bitstream-charter]{mathdesign}
\DeclareSymbolFont{usualmathcal}{OMS}{cmsy}{m}{n}
\DeclareSymbolFontAlphabet{\mathcal}{usualmathcal}

\def\be{\begin{equation}}
\def\ee{\end{equation}}
\def\ba{\begin{eqnarray}}
\def\ea{\end{eqnarray}}

\def \pmbtext#1{\leavevmode
     \setbox0\hbox{#1}
     \kern0,4pt \copy0 \kern-\wd0
     \kern-0,2pt \raise0,3pt \box0 }

\def\kk{{\bf k}}
\def\pp{{\bf p}}
\def\qq{{\bf q}}

\def\rr{{\bf r}}
\def\RR{{\mathbb{R}}}
\def\HH{{\cal H}}

\def\kpn{k_{\perp}}
\def\ppn{p_{\perp}}
\def\qpn{q_{\perp}}

\newfam\msbmfam
\font\tenmsbm=msbm10 \textfont\msbmfam=\tenmsbm

\def \smalloiintaux{\int\mkern-12mu\circ\mkern-12mu }
\def \bigoiintaux{\int\mkern-16mu\bigcirc\mkern-17mu }
\def \oiint{\mathchoice{\bigoiintaux}{\smalloiintaux}{\smalloiintaux}{\smalloiintaux}\int}

\begin{document}

\begin{center}
{\Large \textbf{Wave turbulence \\
A solvable problem applied to Navier-Stokes}}
\end{center}

\begin{center}
S\'ebastien Galtier
\end{center}

\begin{center}
Universit\'e Paris-Saclay\\
Laboratoire de Physique des Plasmas, École polytechnique, 91128 Palaiseau, France\\
{\small \sf sebastien.galtier@lpp.polytechnique.fr}
\end{center}

\begin{center}
\today
\end{center}

% For convenience during refereeing (optional), you can turn on line numbers by uncommenting the next line:
%\linenumbers

\section*{Abstract}
{\bf 
Wave turbulence and eddy turbulence are the two regimes that we may encounter in nature. The attention of fluid mechanics being mainly focused on incompressible hydrodynamics, it is usually the second regime that is treated in books, whereas waves are often present in geophysics and astrophysics. In these lecture notes, I present the theory of wave turbulence which is free from the closure problem encountered in eddy turbulence. Basically, the wave amplitude is introduced in a multiple time scale method as a small parameter to derive the so-called kinetic equations from which exact results can be obtained (power-law spectra, direction of the cascade, Kolmogorov's constant) and compared with the data. Two hydrodynamic applications are considered with capillary waves and inertial waves, the first leading to isotropic turbulence and the second to anisotropic turbulence.}

\vspace{10pt}
\noindent\rule{\textwidth}{1pt}
\tableofcontents\thispagestyle{fancy}
\noindent\rule{\textwidth}{1pt}
\vspace{10pt}

%%%%%%%%%%%%%%%%%%%%%%%%%%%%%%%%%%%%%%%%%%%%%%%%%%%%%%
\begin{figure}[ht]
\center
\centerline{\includegraphics[width=.8\linewidth]{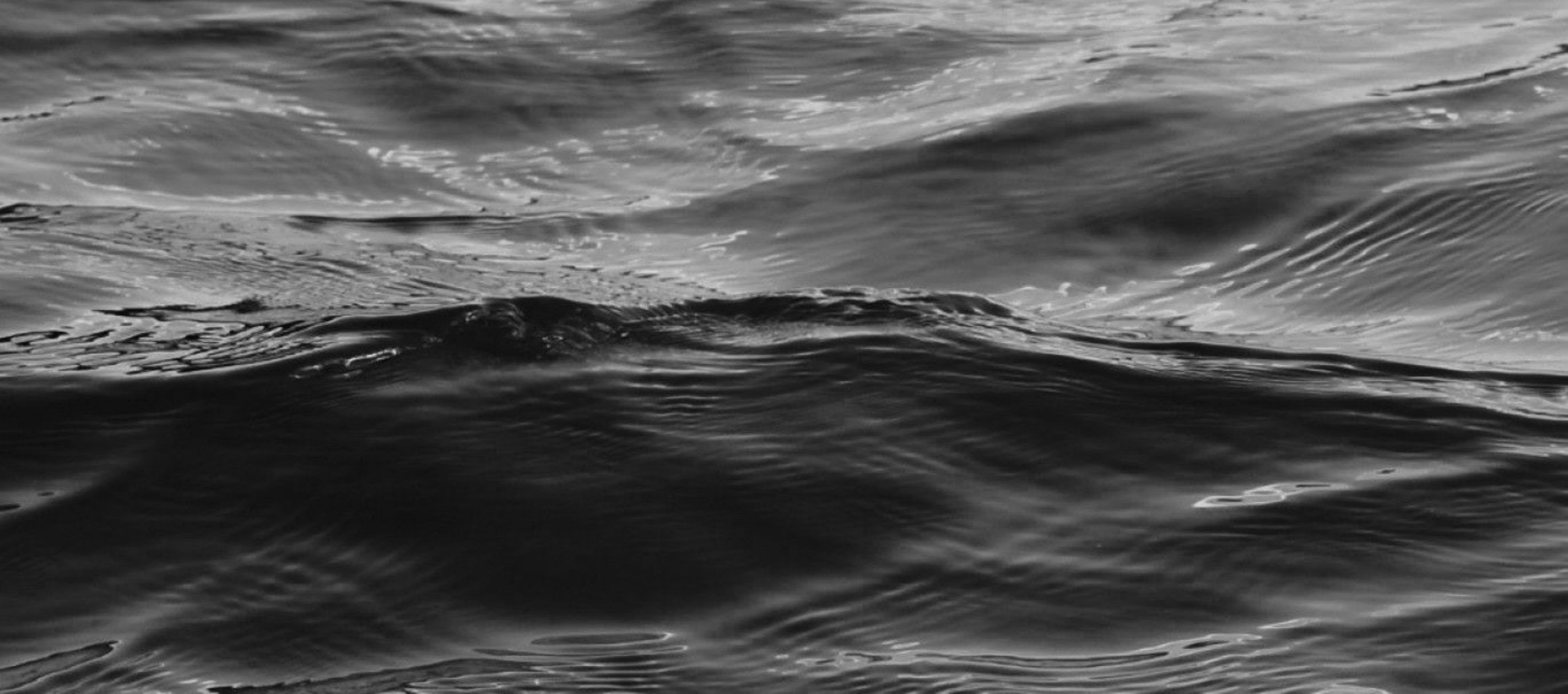}}
\caption{Gravity and capillary wave turbulence.}
\label{Fig0}
\end{figure}
%%%%%%%%%%%%%%%%%%%%%%%%%%%%%%%%%%%%%%%%%%%%%%%%%%%%%%

Wave turbulence offers the possibility of a deep understanding of physical systems composed of an ensemble of random waves interacting nonlinearly. The reason for this is, first of all, the possibility of analytically deriving a set of integro-differential equations for spectral cumulants -- the so-called kinetic equations -- which are free from the closure problem classically encountered in eddy turbulence. To achieve this natural asymptotic closure, the wave amplitude is used as a small parameter. Second, exact solutions can be found from the kinetic equations. In addition to the usual thermodynamic solutions, the kinetic equations have finite flux solutions that capture the flow of conserved densities from sources to sinks. Third, these exact solutions correspond to power-law spectra that can be compared to the data. The number of experiments, observations, and diagnostics has increased dramatically over the past two decades, and now, with direct numerical simulations, wave turbulence has become a leading field in turbulence from which new fundamental questions have arisen \cite{GaltierCUP2023}. 

The Navier-Stokes equations are fundamental in turbulence because it is from them and from laboratory experiments based on water or air that the first theoretical advances (concepts, exact laws) have been made. This is a reason why the literature on turbulence focuses mainly -- if not entirely -- on incompressible hydrodynamics, thus excluding a whole series of applications where waves make a major contribution to the dynamics. Yet turbulent systems containing waves are very numerous, and incompressible hydrodynamics (pure eddy turbulence) seems to be quite singular. In essence, wave turbulence  concerns only systems in which waves can be excited, which excludes a priori the classical Navier-Stokes equations, the archetypal system of turbulence. However, wave turbulence can also be found in incompressible hydrodynamics if one modifies this system to include, for example, surface (gravity and capillary) waves, or if one adds the Coriolis force which introduces inertial waves. 

A priori, we can distinguish two regimes in wave turbulence: the one where the waves are of weak amplitude and the one where they are not. The regime that concerns us here is the first case called weak wave turbulence. I will follow the usual practice and simply write 'wave turbulence' to avoid thinking that weak (wave) turbulence also means small Reynolds number: indeed, as for eddy turbulence, the Reynolds number is not limited or, in other words, it is only limited by the effect of viscosity and the size of the experiment. 
The existence of a small parameter -- the amplitude of the wave -- allows a systematic approach to the problem. The general idea is that the weak amplitude of the waves makes it possible to distinguish the temporal evolution of the amplitude from that of the phase, the first evolving more slowly than the second. This separation of the time scales can be formally written using a multiple time scale method. This is the condition applied to obtain a uniform asymptotic development which leads to a natural closure of the hierarchy of equations and finally to the so-called kinetic equations. 

In these lecture notes, I will introduce the multiple time scale method to an archetypal wave equation that involves three-wave interactions. I will show in detail how the kinetic equation can emerge from this development. This is a technical part that has rarely (if never) been described in detail since the seminal paper \cite{Benney1966}. 
A brief history will allow us to understand how the theory of wave turbulence was built. For pedagogical reasons, the multiple time scale method will first be applied to a simple example: the Duffing model. We will see that the main problem is the existence of secular terms proportional to time which can break the uniformity of the development. This is precisely the condition applied to eliminate them that leads to the kinetic equations of wave turbulence. I will examine two examples from the modified Navier-Stokes equations: capillary wave turbulence and inertial wave turbulence involving three-wave interactions. The first  implies isotropy while the second leads to anisotropy, and thus to a new difficulty. Finally, using a toy model, I will show how the Kolmogogov-Zakharov spectrum can be derived.

%%%%%%%%%%%%%%
\section{Introduction to wave turbulence: origin and applications}
%%%%%%%%%%%%%%
It is essentially in the field of oceanography and the study of surface waves that we find the beginnings of the theory of wave turbulence \cite{Phillips1981}. This work began in the late 1950s, at a time when the surface waves were treated essentially in a linear manner, the spectra as a superposition of linear waves and non-linear effects restricted, for example, to periodic wave distortion phenomena described for a long time by Stokes \cite{Stokes1847}. 
There are two kinds of surface waves: gravity waves -- that is, waves on the ocean (see Figure \ref{Fig0}), and at smaller scale capillary waves. The Navier-Stokes equations, modified by the force of gravity, are the theoretical framework from which the first results emerged. To simplify the problem, the movements are generally assumed to be irrotational: this corresponds to an air-water interface disturbed by a unidirectional blowing wind -- a typical condition encountered in the open sea. The problem is then reduced to Bernoulli's equation applied to the free surface of the fluid to which is added a Lagrangian equation to describe the deformation of the fluid surface and another to take into account the additional hypothesis called deep water. 

%%%%%%%%%%%%%%%%%%
\subsection{Resonant wave interactions}
%%%%%%%%%%%%%%%%%%

Work in the late 1950s and early 1960s led to a major theoretical breakthrough with the discovery of the existence of resonant interactions\footnote{Note that earlier studies on three- and four-wave resonance had already been carried out, for example, in the field of thermal conduction in crystals \cite{Peierls1929}. See also \cite{Nordheim1928} who wrote the first kinetic equation: it was for four-wave interactions in the context of the electron theory of conductivity.} between non-linear waves of weak amplitude. The waves in question are of course gravity waves. The idea has its origin in the work of \cite{Phillips1960}, \cite{Longuet1962} and \cite{Hasselmann1962}\footnote{K. Hasselmann was awarded the Nobel prize in Physics in 2021 'for the physical modelling of Earth's climate, quantifying variability and reliably predicting global warming'.} who tried to understand how the non-linear interactions between gravity waves could redistribute the energy initially present in these modes. The underlying idea is that in the initial phase of the development of turbulence it is the resonant interactions that provide the dominant mechanism for transferring energy from one wave to another. The ultimate phase of development had already been studied previously by \cite{Phillips1958} who, using dimensional analysis, was able to make a spectral prediction for strong (wave) turbulence which is still in use today (the so-called critical balance). 

After long calculations based on a classical perturbative development, it was possible to demonstrate that the energy redistribution was generally negligible for interactions involving less than four waves, but that it became important for four waves if the following resonance relationship was satisfied
\be \label{R1L1}
\left\{
    \begin{array}{ll}
        \kk_1 + \kk_2 = \kk_3 + \kk \, ,   \\ 
        \omega_1 + \omega_2 = \omega_3 + \omega \, , 
    \end{array}
\right.
\ee
with the dispersion relation $\omega^{2}=g\vert {\bf k} \vert$ ($g$ being the acceleration of gravity). The experimental demonstration of the existence of resonant interactions was then published by \cite{Longuet1966} and \cite{McGoldrick1966}. Although the existence of resonant wave interactions appeared to be real, their interpretation was then subject to debate because the energy exchange between resonant waves also led to a linear growth (in time) -- the secular terms -- of the amplitude of the waves ultimately breaking the hypothesis underlying the existence of resonant interactions, i.e. the presence of waves of weak amplitude; in this case, the perturbation is said to be non-uniform. It was therefore not clear whether this energy transfer could be really effective. 

An answer to this question was first proposed statistically by \cite{Hasselmann1962} who considered a random set of gravity waves. The hypothesis of weak non-linearities was then used to justify another hypothesis, that of a Gaussian statistics which greatly simplifies the calculations by eliminating in particular moments of odd order (see also \cite{Drummond1962}).

%%%%%%%%%%%%%%%%%%
\subsection{Multiple time scale method}
%%%%%%%%%%%%%%%%%%

A second major theoretical breakthrough came with Benney's work which showed that the problem of wave turbulence is similar to that of weakly coupled oscillators in mechanics \cite{2020Akylas}. With a new mathematical technique for the analysis of dispersive wave packets involving two time scales, \cite{Benney1962} obtained a relatively simple form for the (discrete and non-statistical) equations which govern the temporal evolution of resonant modes. In this approach, the secular terms disappear: they are somehow absorbed by the slow variation of the wave amplitude. This work shows, in a simple way, how the exchange of energy between four gravity waves is carried out: this exchange takes place while conserving energy. Here we recover a property of strong turbulence, that of detailed conservation. 

This approach paved the way for the use of a new mathematical technique called multiple time scale method \cite{Poincare1893,Sturrock1957,Nayfeh2004}, which made it possible to demonstrate that the (continuous) statistical equations of wave turbulence have a natural closure due to the separation of scales in time. As shown by \cite{Benney1966} for quadratic non-linearities and \cite{Benney1967} for cubic non-linearities, the equations of wave turbulence are asymptotically valid over long periods of time and do not require the statistical assumption of Gaussianity made by \cite{Hasselmann1962}. The method of multiple time scales offers a systematic and consistent theoretical framework in which the procedure expansion allows in principle to determine the slow rate of change of the amplitude of the waves at any order in $\epsilon$ \cite{Benney1967b,Benney1969}. The so-called kinetic equation of gravity wave turbulence then takes the following (schematic) integro-differential form in three-dimension
\ba \label{EqKin}
\frac{\partial N(\kk)}{\partial t} &=&  \epsilon^4 \int_{\mathbb{R}^{9}} S(\kk,\kk_{1},\kk_{2},\kk_{3}) [N_1 N_2 (N+N_3) - N_3 N (N_1+N_2)]  \delta(\omega_1 + \omega_2 - \omega_3 - \omega) \nonumber \\
&& \delta(\kk_1 + \kk_2 - \kk_3 -\kk)  d\kk_1 d\kk_2 d\kk_3 \, ,
\ea
with $N \equiv N(\kk)=E(\kk)/\omega$ the wave action spectrum\footnote{By analogy with plasma physics, wave action is often associated with particles. This is a quantity that is sometimes conserved in four-wave processes. This is the case for gravity wave turbulence.} and $E(\kk)$ the energy spectrum. The Dirac functions translate the resonance condition (\ref{R1L1}) discussed above. The presence of the $\epsilon^4$ factor means that the time scale (normalized to the wave period $\sim 1 / \omega$) on which the spectra are modified by the non-linear dynamics is of the order of $\mathcal{O} (1/\epsilon^{4})$.  Therefore, it is a relatively slow process. 

%%%%%%%%%%%%%%%%%%
\subsection{Kolmogorov-Zakharov spectrum}
%%%%%%%%%%%%%%%%%%

Parallel to the work carried out in the Western world, major theoretical advances were also made in the East. From the beginning of the 1960s, the Soviet school became interested in wave turbulence mainly through plasma physics \cite{Sagdeev1966,Vedenov67} from which certain notations and vocabulary were borrowed (one speaks, for example, of kinetic equation or collision integral). In passing, it is curious to note that this work was carried out simultaneously by the two parts of the world without much communication between them. In particular, by the assumption known as random phase approximation, kinetic equations of wave turbulence were proposed in a primitive form by \cite{Kadomtsev1963} for a problem of plasma physics, then in a modern form by Zakharov and his collaborators \cite{Zakharov1965,Zakharov1966,Zakharov1967,Zakharov1967b}. This work is generally based on a Hamiltonian approach to the problem, whereas it is the Eulerian approach which was mainly used in the West. 

The random phase approximation leads, in practice, to the same kinetic equations as by the method of multiple time scales where the random phase occurs naturally dynamically (and therefore is not assumed). From these integro-differential equations, a major breakthrough was achieved with the discovery of a conformal transformation to extract from the non-linear kinetic equations the exact power law solutions. This transformation -- now called the Zakharov transformation\footnote{Zakharov was a student of Sagdeev. He defended his PhD thesis on surface waves in 1966 with several fundamental results in wave turbulence to his credit, such as the discovery of exact solutions to the kinetic equations. This discovery is reported in the article by \cite{Zakharov1965} in which the author first verified that the collision integral (obtained from a relatively simple ad hoc model of three-wave interactions) tended towards $\pm \infty$ for two power law exponents. He then demonstrated that the solution (the energy spectrum) associated with the index exactly in the middle of this interval cancels non-trivially the collision integral: Zakharov had just discovered an exact stationary solution. The discovery of the so-called Zakharov transformation came shortly after \cite{Zakharov1966}.} -- was first proposed for capillary wave turbulence involving triadic interactions \cite{Zakharov1966,Zakharov1967}, then for Langmuir wave turbulence where the interactions are quartic \cite{Zakharov1967b,Kaner1970}. There are two types of solution: the zero flux solution (which was the regime studied by \cite{Nordheim1928}) and the non-zero constant flux solution. The first case corresponds to the thermodynamic solution (constant entropy) and the second to the Kolmogorov-Zakharov spectrum, which is the most interesting solution because it is non-trivial and corresponds to a cascade.

%%%%%%%%%%%%%%%%%%%%%%%%%%%%%
\subsection{Applications of wave turbulence}
%%%%%%%%%%%%%%%%%%%%%%%%%%%%%

There are many examples of application of wave turbulence. Below is a non-exhaustive list of applications with some references.  

\medskip

\noindent
$\bullet$ {\bf Surface wave:} These are the first applications of wave turbulence. They include capillary waves and gravity waves. The former involve three-wave interactions, the theory of which was published in English by \cite{Zakharov1967} in the deep water limit. Note that the shallow water limit is also the subject of studies (see for example \cite{Clark2014}). Gravity wave turbulence is a problem that needs to be addressed at the level of four waves \cite{Hasselmann1962,BenneyNewell1967}. This regime has been well reproduced in the laboratory or by direct numerical simulations (see eg. \cite{Deike2011,Zhang2022}). Its detection at sea is more difficult but not impossible (see eg. \cite{Hwang2000,Lenain2017}). The two subjects being linked, several experiments deal with the interaction between gravity and capillary waves. All in all, it is a subject that is still very much under study (see the review by \cite{Falcon2022}). 

\medskip

\noindent
$\bullet$ {\bf Internal gravity wave:} These waves are a variant of the previous ones in the sense that we are interested here in gravity waves in oceans. These waves contribute dynamically to the turbulent transport of heat, which is important to understand to properly evaluate the impact of  oceans on the climate \cite{MacKinnon2017}. Internal gravity wave turbulence is an anisotropic three-wave problem for which an Eulerian theory has been developed by \cite{Caillol2000} as well as several experiments (see eg. \cite{Savaro2020}). 

\medskip

\noindent
$\bullet$ {\bf Inertial wave:} This is the closest example to the standard case of eddy turbulence in the sense that the equations are those of Navier-Stokes which are modified simply by adding the Coriolis force. Inertial wave turbulence is a three-wave anisotropic problem whose theory has been published by \cite{Galtier2003}. Several experiments have been made to study this regime (see e.g. \cite{Monsalve2020}). This example will be discussed later. 

\medskip

\noindent
$\bullet$ {\bf Rossby wave:} These waves appear in a situation of differential rotation. They are used in the modeling of planetary atmospheres, the interiors of gaseous planets and stars; these are often referred to as planetary waves \cite{LonguetH1967}. In wave turbulence, the dynamics is driven by three-wave processes with a dominance of non-local interactions \cite{Balk1990,Balk1990b}. 

\medskip

\noindent
$\bullet$ {\bf Plasma wave:} As mentioned above, we find the beginnings of wave turbulence in the field of plasma physics \cite{Sagdeev1966,Vedenov67}. In this very vast domain, waves are legion. Incompressible magnetohydrodynamics (MHD) is an important example of application because it is a model often used in astrophysics. We talk about Alfv\'en wave turbulence which involves three-wave interactions. The theory was published by \cite{Galtier2000}: it is a case where the anisotropy is so strong that the cascade is completely inhibited in the direction of the strong applied magnetic field. 

\medskip

\noindent
$\bullet$ {\bf Geodynamo wave:} Geodynamo waves are defined as waves present in the Earth's outer liquid core. These waves take part in the dynamo effect, i.e. the physical mechanism that maintains the Earth magnetic field \cite{Finlay2008}. These are magnetostrophic waves and inertial waves. The three-wave theory models a homogeneous medium with a small Rossby number \cite{Galtier2014}: in this framework, an anisotropic inverse cascade of hybrid helicity is predicted, which could be at the origin of the regeneration of the magnetic field on a large scale \cite{Menu2019}. 

\medskip

\noindent
$\bullet$ {\bf Acoustic wave:} Acoustic wave turbulence is a priori driven by three-wave interactions but these waves are semi-dispersive. This is a critical situation for the application of wave turbulence because the uniformity of the development is not guaranteed. The first works on the subject date back to the early 1970s \cite{Zakharov1970}. The way in which the asymptotic is subtly modified is discussed by \cite{Newell1971}, then by \cite{Lvov1997}. It can be shown for this regime that energy is at best redistributed according to rays (in three-dimensional Fourier space). Note that a similar situation can be found in compressible MHD \cite{Galtier2023}.

\medskip

\noindent
$\bullet$ {\bf Elastic wave:} It is a turbulence produced by a thin elastic plate or by the introduction of long polymer molecules in a 
liquid \cite{Steinberg2021}. In the former case, we are in a very different situation from the traditional turbulence produced by a fluid. The vibrations of a plate can in theory produce weak or strong wave turbulence, depending on the forcing. The theory of wave turbulence has been published by \cite{During2006}: it is a four wave problem characterized by a direct energy cascade. Although the wave action is not conserved in this problem, direct numerical simulations show an inverse cascade which seems to be established in an explosive way \cite{During2015}. Laboratory experiments using steel plates have been carried out in order to reproduce with varied success the theoretical predictions
\cite{Boudaoud2008,Mordant2008,Cobelli2009,Mordant2010}. 

\medskip

\noindent
$\bullet$ {\bf Optical wave:} Wave turbulence is also found in the field of non-linear optics. The multi-dimensional non-linear Schr\"odinger equation can be used to describe the evolution of quasi-monochromatic plane wave envelopes \cite{Sulem1999}. The theory is explained by \cite{Dyachenko1992}: this wave turbulence is governed by four-wave interactions and an intermittency mechanism characterized by a collapse phenomenon in physical space can occur. In the case of a reduction of the problem to one dimension, it can be shown that the dominant resonant interactions imply six waves, involving a much slower non-linear dynamics \cite{Laurie2012}. All this work is part of the study of non-linear effects on the propagation of incoherent optical beams \cite{Mitchell1996,Picozzi2014}. 

\medskip

\noindent
$\bullet$ {\bf Quantum turbulence:} This subject is very close to the previous one in the sense that the model used is a variant of the non-linear Schr\"odinger equation: by changing the sign of the non-linear term, the interaction becomes repulsive and the physics of turbulence is modified. The associated equation -- also called the Gross-Pitaevskii equation -- describes a Bose gas at very low temperatures. Note that the emergence of quantum turbulence in an oscillating Bose-Enstein condensate has been experimentally demonstrated \cite{Henn2009}. The wave turbulence regime is described by \cite{Dyachenko1992}: it is shown that four-wave interactions conserve energy and wave action. The inverse cascade associated with the latter leads to the formation of a stable condensate, i.e. the accumulation of wave action in the $k=0$ mode. Direct numerical simulations in two \cite{Nazarenko2006BEC,Nazarenko2007} and three  \cite{Proment2009,Proment2012} dimensions illustrate this regime. 

\medskip

\noindent
$\bullet$ {\bf Kelvin wave:} Kelvin waves are studied in the context of superfluids whose temperature is close to the absolute zero. These waves can propagate along filaments of vorticity and modify the dynamics of turbulence \cite{Vinen2000,Kivotides2001}. Kelvin's theory of wave turbulence has been proposed by \cite{Kozik2004}: it involves six-wave processes that conserve both energy and wave action. As these cascade processes are extremely slow, it is particularly interesting to consider a local model of non-linear diffusion \cite{Nazarenko2006}. 

\medskip

\noindent
$\bullet$ {\bf Gravitational wave:} This example illustrates the wide range of possible applications of wave turbulence. It is about cosmology and more precisely the birth of the Universe: primordial gravitational waves could be at the origin of the inflation phase of the Universe, a question which is still open to this day and which touches the limits of our knowledge \cite{Galtier2020gw}. The theory has been published by \cite{Galtier2017,Galtier2021}: the dynamics is governed by four-wave processes sufficiently symmetric to conserve wave action. The latter is characterized by an explosive inverse cascade. It is interesting to note that this regime is close to that of elastic waves in the strong tension limit as shown by the theoretical, numerical and experimental work of \cite{Hassaini2019}. 

%%%%%%%%%%%%%%%%%%%%%%%%%%%%%%
\section{Duffing's equation: a pedagogical model}
\subsection{Classical perturbation method}
%%%%%%%%%%%%%%%%%%%%%%%%%%%%%%
The aim of this section is to discuss on a pedagogical model the idea of the multiple time scale method through a toy model, the Duffing equation 
\be
\frac{d^2 f}{dt^2} + f = -\epsilon f^3 \, ,
\ee
where $\epsilon$ is a small parameter ($0<\epsilon\ll 1$) which measures the intensity of the non-linearity and $f$ is a function of time only. The main solution to this equation is a harmonic oscillator. This solution will, however, be slightly modified over time by the presence of the small non-linear perturbation (right hand side term). Let us first use the classical perturbation theory and introduce the following development
\be
f = \sum_{n=0}^{+\infty} \epsilon^{n} f_{n} \, . 
\ee
We then obtain an infinite system of equations which, for the first three orders, can be written
\begin{subequations}
\begin{align}
\mathcal{O} (\epsilon^0):& \quad \frac{d^2 f_{0}}{dt^2} + f_{0} = 0 \, , \\
\mathcal{O} (\epsilon^1):& \quad \frac{d^2 f_{1}}{dt^2} + f_{1} = -f^{3}_{0} \, , \\
\mathcal{O} (\epsilon^2):& \quad \frac{d^2 f_{2}}{dt^2} + f_{2} = -3f^{2}_{0}f_{1} \, .
\end{align}
\end{subequations}
We can see that the solution to a given order will affect the solution to the higher order. The first solution is trivially $f_{0}=A \cos(t+\phi)$. By introducing this one in the second equation, we can deduce the exact solution for $f_{1}$. If we stop at this order, the solution reads
\be
f(t)= A \cos(t+\phi) + \epsilon A^{3} \left[ -\frac{3}{8} t \sin (t+\phi) + \frac{1}{32} \cos (3t+3\phi) \right] + \mathcal{O} (\epsilon^2) \, .
\label{19}
\ee
As a first approximation, this system behaves well as a harmonic oscillator. The condition to be checked is that the non-linear perturbation remains of weak amplitude and the time considered relatively short. On the other hand, over long periods of the order of $\mathcal{O} (1/\epsilon)$, this solution is modified in a non-negligible way. The term which is at the origin of it, in $\epsilon t$, is said to be secular. For even longer periods of time the development diverges; this divergence is all the more so as certain secular terms of a higher order reinforce it. Then, the uniformity of the development is broken. 

We can pursue the analysis to evaluate the period of oscillations. To do this, the Duffing equation, which has been previously multiplied by the derivative of $f$, is integrated
\be
\frac{1}{2} \left(\frac{df}{dt}\right)^{2} + \frac{1}{2}f^{2} + \frac{1}{4}\epsilon f^4 = E \, , 
\ee
where $E$ is a constant. One obtains
\be
dt = \frac{df}{\sqrt{2E-f^{2} - \frac{1}{2}\epsilon f^{4}}} \, .
\ee
This equation can be integrated over a period $T$. By posing $f=a \sin \theta$, we obtain the following expression which is exact in its integral form
\be
T = 4 \int_{0}^{\pi/2} \frac{d \theta}{\sqrt{1+\epsilon a^{2} (1-\frac{1}{2}\cos^{2} \theta)}} 
= 2 \pi \left( 1 - \frac{3}{8} \epsilon a^{2} + \mathcal{O} (\epsilon^2 a^{4}) \right) \, .
\ee
A Taylor expansion is used to evaluate the integral. It is found that the $2\pi$ period of the harmonic oscillator is corrected by the non-linear perturbation. The smaller the non-linearity, the smaller the correction. 

What information is useful for our problem? This very simple system illustrates schematically the problems that we can encounter in a classical perturbation  development. Such a development should allow us, in principle, to follow the evolution of the dynamics of wave turbulence at various orders in $\epsilon$. The main problem is to ensure that the development remains neat (or uniform). In other words, this means making sure that no secular term of the order $n$ comes to interfere in the dynamics at an order $m$, such as $m < n$, in order not to find oneself in the same situation as \cite{Hasselmann1962}, i.e. in the presence of a secular term. This type of term constitutes an obstacle to the closure of the hierarchy of equations. The more sophisticated method of multiple time scales \cite{Poincare1893} will allow us to solve this problem. 

%%%%%%%%%%%%%%%%%%%%%%%%%%%%%%
\subsection{Multiple time scale method}
%%%%%%%%%%%%%%%%%%%%%%%%%%%%%%

Let us take the Duffing equation again and introduce the following independent time variables
\be
T_{n} \equiv \epsilon^{n} t \quad \text{with} \quad n=0, 1, 2, ... \, .
\label{METM1}
\ee
The power series development of the variable $f$ is then written ($t \equiv T_0$)
\be
f = \sum_{n=0}^{+\infty} \epsilon^{n} f_{n}(t, T_{1}, T_{2},...) \, ,
\label{METM2}
\ee
and the time derivative becomes a sum of partial derivatives
\be
\frac{d}{dt} = \sum_{n=0}^{+\infty} \epsilon^{n} \frac{\partial}{\partial T_{n}} \, .
\label{METM3}
\ee
The introduction of relations (\ref{METM1})--(\ref{METM3}) into the Duffing equation gives
\ba
&&\left(\frac{\partial}{\partial t} + \epsilon \frac{\partial}{\partial T_{1}} + \epsilon^{2} \frac{\partial}{\partial T_{2}} + ...\right)^{2} (f_0 + \epsilon f_1 + \epsilon^{2} f_{2} +...)  + (f_0 + \epsilon f_1 + \epsilon^{2} f_{2} +...) = \nonumber \\
&& -\epsilon (f_0 + \epsilon f_1 + \epsilon^{2} f_{2} +...)^{3} \, . 
\ea
We obtain an infinite system of equations which, for the first two orders, can be written
\begin{subequations}
\begin{align}
\mathcal{O} (\epsilon^0):& \quad \frac{\partial^2 f_{0}}{\partial t^2} + f_{0} = 0 \, , \\
\mathcal{O} (\epsilon^1):& \quad \frac{\partial^2 f_{1}}{\partial t^2} + f_{1} + 2 \frac{\partial^{2} f_{0}}{\partial t \partial T_{1}}= -f^{3}_{0} \, . \label{118}
\end{align}
\end{subequations}
We can see the presence of a new term in the time evolution equation of $f_{1}$ whose importance emerges over long times. We look for a solution of the form
\be
f_{0}=A(T_{1},T_{2},...) \cos(t+\phi(T_{1},T_{2},...)) \, ,
\ee
with an amplitude $A$ and a phase $\phi$ which can vary slowly over time (hence the dependence in $T_1$, $T_2$, etc). 
The introduction of previous expression in (\ref{118}) gives the relation
\ba
\frac{\partial^2 f_{1}}{\partial t^2}  + f_{1} &=& 2 \frac{\partial A}{\partial T_{1}} \sin(t+\phi) 
+ \left(2 A \frac{\partial \phi}{\partial T_{1}} - \frac{3}{4} A^{3}\right) \cos(t+\phi) -\frac{A^{3}}{4} \cos(3(t+\phi)) . 
\ea
To ensure that the $f_{1}$ solution does not include a secular term, the following conditions must be imposed
\be
\frac{\partial A}{\partial T_{1}} = 0 \quad \text{and} \quad \frac{\partial \phi}{\partial T_{1}} = \frac{3}{8} A^{2} \, ,
\ee
which gives $\phi=(3/8)A^{2}\epsilon t + \theta$. The solution is then written
\be \label{duffing2}
f(t) = A \cos\left(t + \frac{3}{8} A^2 \epsilon t + \theta\right) + 
\frac{\epsilon A^3}{32} \cos\left(3t + \frac{9}{8} A^2 \epsilon t + 3 \theta\right) + \mathcal{O} (\epsilon^2) \, , 
\ee
with $A(T_{2},T_{3},...)$ and $\theta(T_{2},T_{3},...)$. In the order of the truncation, we can therefore consider that $A$ and $\theta$ are constant. 
Note that this development is compatible with the previous one (\ref{19}) for $t \le \mathcal{O} (\epsilon)$. This example illustrates the fact that we can achieve a uniform development in a systematic way using the multiple time scale method. 

%%%%%%%%%%%%%%%%%%%%%%%%%%%%%%%%%%%%%%%%%%%%%%%%%%%%%%
\begin{figure}[ht]
\center
\centerline{\includegraphics[width=.8\linewidth]{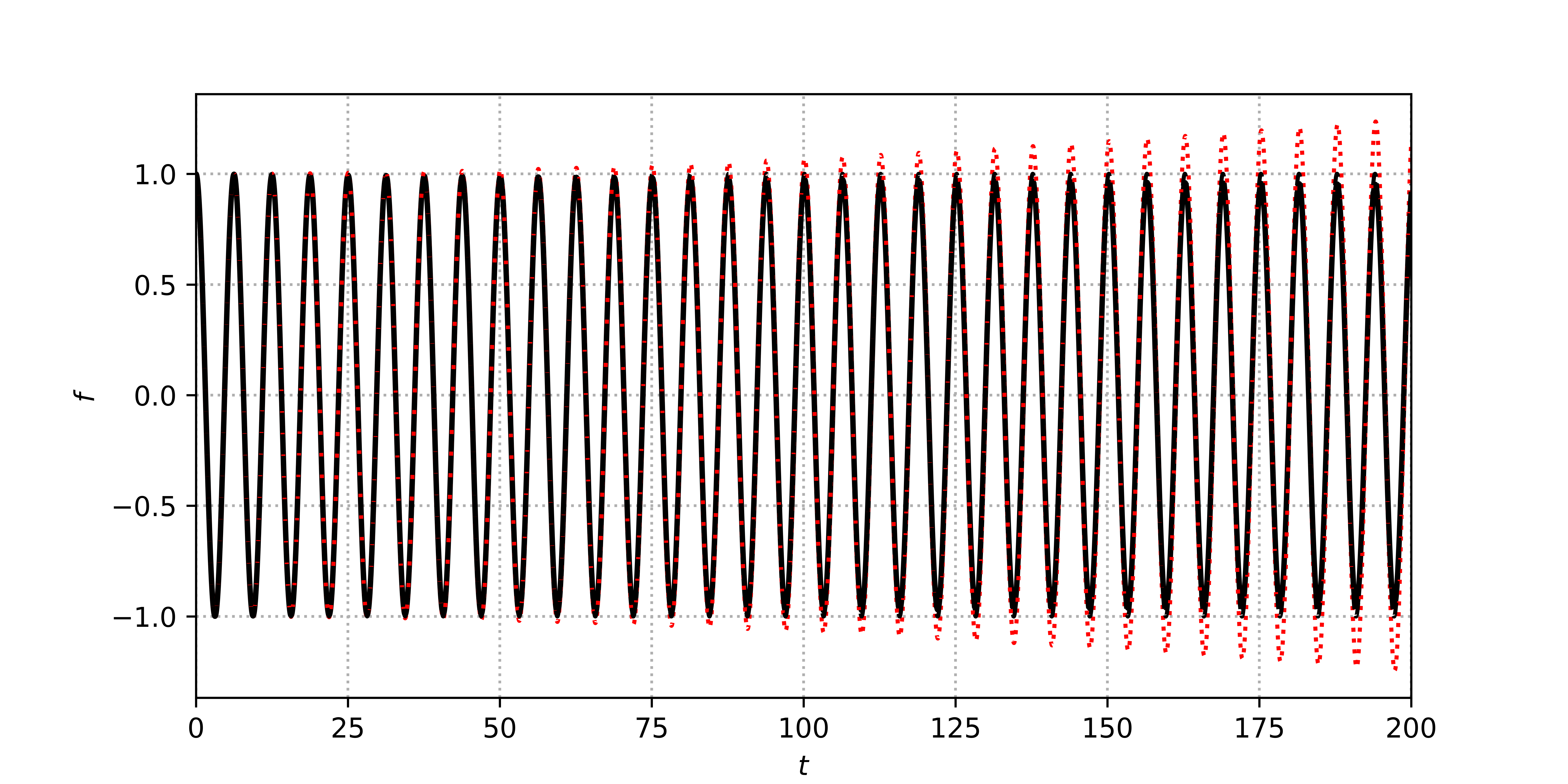}}
\centerline{\includegraphics[width=.8\linewidth]{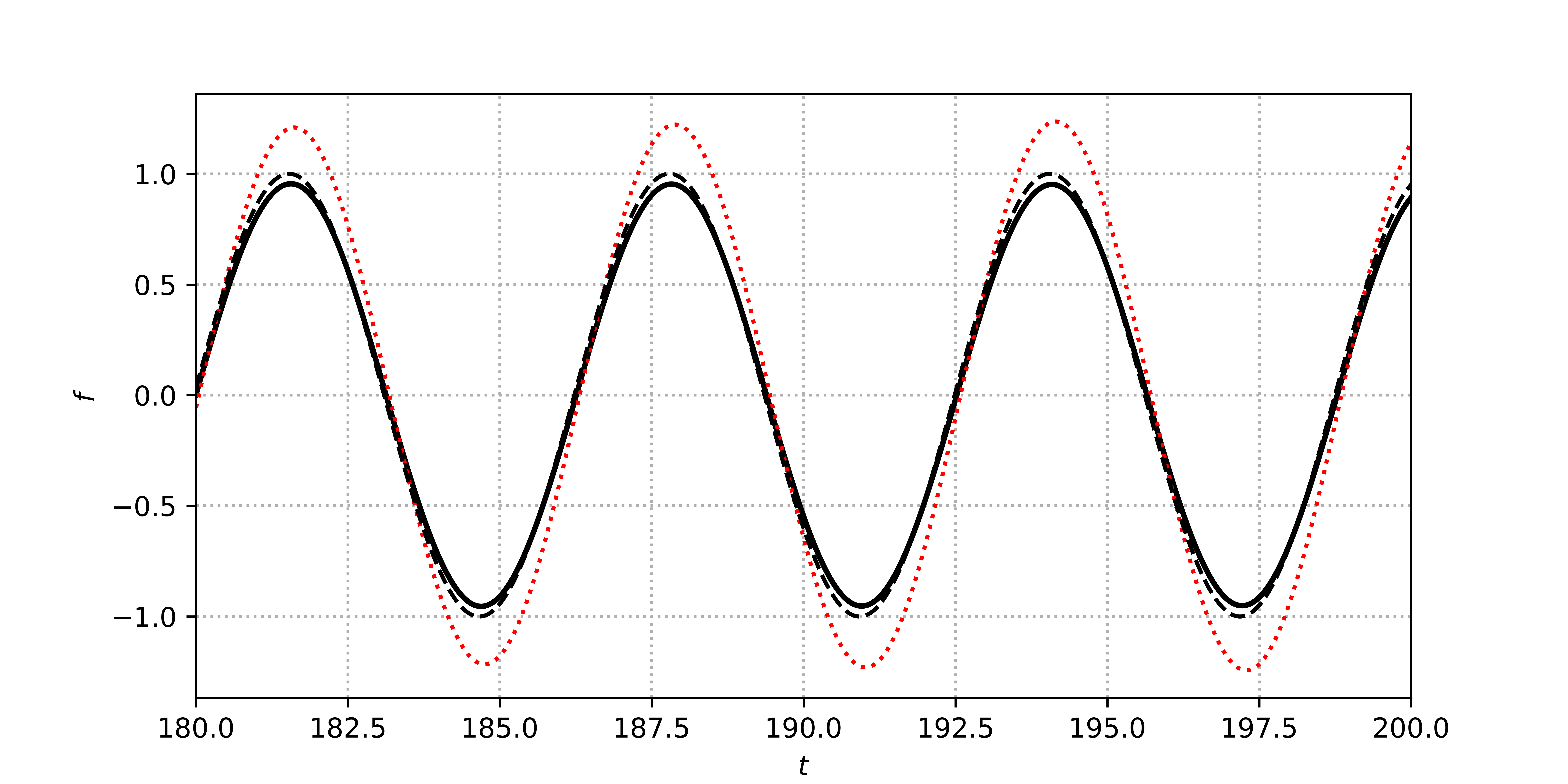}}
\caption{Variation of $f(t)$ (solid line) obtained by a numerical simulation of the Duffing equation with $\epsilon=0.01$. The solutions (\ref{19}) and (\ref{duffing2}) are plotted in red dotted and black dashed lines, respectively.}
\label{Fig6-1ab}
\end{figure}
%%%%%%%%%%%%%%%%%%%%%%%%%%%%%%%%%%%%%%%%%%%%%%%%%%%%%%

In Figure \ref{Fig6-1ab} we show the result of a numerical simulation of the Duffing equation with $\epsilon=0.01$. The function $f(t)$ is plotted in solid line. At the top, we see the evolution for $t \in [0,200]$ and at the bottom, we display an enlargement of the end of the simulation. The mathematical solutions (\ref{19}) and (\ref{duffing2}) are plotted in dotted and dashed lines, respectively. The divergence with $t$ of the solution (\ref{19}) whose origin is the secular term can be seen. As expected, this divergence occurs over a time $t$ of the order of $\mathcal{O} (1/\epsilon)$ (the reference time is the wave period). On the other hand, the solution (\ref{duffing2}) remains close to the real solution until the final time.

%%%%%%%%%%%%%%%%%%%%%%%%%%%%%%
\section{Sequential time closures for three-wave interactions}
%%%%%%%%%%%%%%%%%%%%%%%%%%%%%%
\subsection{Fundamental equation}

We present here a detailed derivation of the kinetic equation for three-wave interactions, a situation of interest for the Navier-Stokes equations when capillary waves or inertial waves are involved, two examples that will be used to illustrate the theory of wave turbulence. 

Consider the following inviscid model
\be
{\partial u({\bf x},t) \over \partial t} = {\cal L}(u) + \epsilon \, {\cal N}(u,u) \, ,
\label{eq1}
\ee
where $u$ is a random variable such that $\langle u \rangle = 0$ (with $\langle \rangle$ an ensemble average), ${\cal L}$ is a linear operator which guarantees that the waves are linear solutions to the problem, and ${\cal N}$ is a non-linear quadratic operator (thus involving a priori triadic interactions). The coefficient $\epsilon$ is a small parameter ($0 < \epsilon \ll 1$) which measures the amplitude of the non-linearities (it is originally a measure of the wave amplitude). The Fourier transform in d-dimension is 
\be
\hat u(\kk,t) \equiv {1 \over (2\pi)^d} \int_{\mathbb{R}^{d}} u({\bf x},t) e^{- i \kk \cdot {\bf x}} d{\bf x} \, .
\ee
Note that we consider a continuous medium which can lead to mathematical difficulties connected with  infinite dimensional phase spaces. For this reason, it is preferable to assume a variable spatially periodic over a box of finite size $L$. However, in the derivation of the kinetic equation, the limit $L \to +\infty$ is finally taken (before the long time limit, or equivalently the limit $\epsilon \to 0$). As both approaches lead to the same kinetic equation, for simplicity, we anticipate this result and follow the original approach of \cite{Benney1966}. 

The Fourier transform of equation (\ref{eq1}) takes the following schematic form when the canonical variables are introduced
\be \label{21e}
\left( {\partial \over \partial t} + is_k \omega_k \right) A^{s_k} (\kk,t) = 
\epsilon \sum_{s_{p} s_{q}}  \int_{\mathbb{R}^{2d}} \HH_{-\kk \pp \qq} A^{s_p}(\pp,t) A^{s_q}(\qq,t) \delta(\kk-\pp-\qq) d\pp d\qq \, ,
\ee
with $s_i=\pm$ the directional polarity. On the left hand side, we find $\omega_k$ which is fixed by the dispersion relation (we assume $0 < \omega_k \propto \vert \kk \vert^x$ with $x>1$) while on the right hand side $\HH_{\kk \pp \qq}$ is an operator which depends on the shape of the non-linearities and is symmetrical in $\pp$ and $\qq$ (ie. $\HH_{\kk \qq \pp} = \HH_{\kk \pp \qq}$). We recall that the presence of the Dirac function originates in the convolution product, and that the quadratic nature of the non-linearities leads to triadic interactions. 

Equation (\ref{21e}) is simplified by making the following change of variables (the writing is simplified in passing)
\be
A^{s_k}(\kk,t) = a^{s_k}(\kk,t) e^{-i s_k\omega_k t} = a^{s_k}_{k} e^{-i s_k\omega_k t} \, .
\label{phaseamplitude}
\ee
In the interaction representation, we obtain the fundamental equation
\be
{\partial a^{s_k}_{k} \over \partial t} = \epsilon \sum_{s_{p} s_{q}} \int_{\mathbb{R}^{2d}} \HH_{-\kk \pp \qq} a^{s_{p}}_{p} a^{s_{q}}_{q} 
e^{i \Omega_{k,pq}t} \delta_{k,pq} d\pp d\qq \, ,
\label{eq6}
\ee
with by definition $\delta_{k,pq} \equiv \delta(\kk-\pp-\qq)$ and $\Omega_{k,pq} \equiv s_k\omega_k - s_{p}\omega_p -s_{q}\omega_q$. 
Additional properties can be established if first we use the relation $A^{s_k}(\kk)=(A^{-s_k}(-\kk))^*$, with $*$ the complex conjugate: then, we have necessarily $\HH_{\kk \pp \qq} = \HH^*_{-\kk -\pp -\qq}$. As mentioned above, we consider only zero mean fluctuations in the physical space, which means with our notation $a^{s_k}(\kk=0,t) = 0$ (below, we will assume  statistical spatial homogeneity) and therefore, $\HH_{{\bf 0} \pp \qq} = 0$. Finally, if $\HH_{\kk \pp \qq}$ is purely imaginary, we have $\HH^*_{\kk \pp \qq}=-\HH_{\kk \pp \qq}$. This property may be used to simplify the kinetic equations. 

Equation (\ref{eq6}) highlights the temporal evolution of the wave amplitude: this evolution is a priori relatively slow since it induces a non-linear term proportional to $\epsilon$. We will find this property with the multiple time scale method. The presence of the complex exponential is fundamental for the asymptotic closure: since we are interested in the dynamics over a long time with respect to the wave period, the contribution of this exponential is essentially zero. Only some terms will survive: those for which $\Omega_{k,pq}=0$.  With the condition imposed by the Dirac, we obtain the resonance condition
\be \label{cr2425z}
\left\{
    \begin{array}{ll}
       \kk =  \pp+\qq \, , \\
	s_k \omega_k =  s_{p}\omega_p +s_{q}\omega_q \, .
    \end{array}
\right.
\ee

%%%%%%%%%%%%%%%%%%%%%%%%%%%%%%%%%
\subsection{Dispersion relation and resonance}
%%%%%%%%%%%%%%%%%%%%%%%%%%%%%%%%%

\begin{figure}
\center
\centerline{\includegraphics[width=1\linewidth]{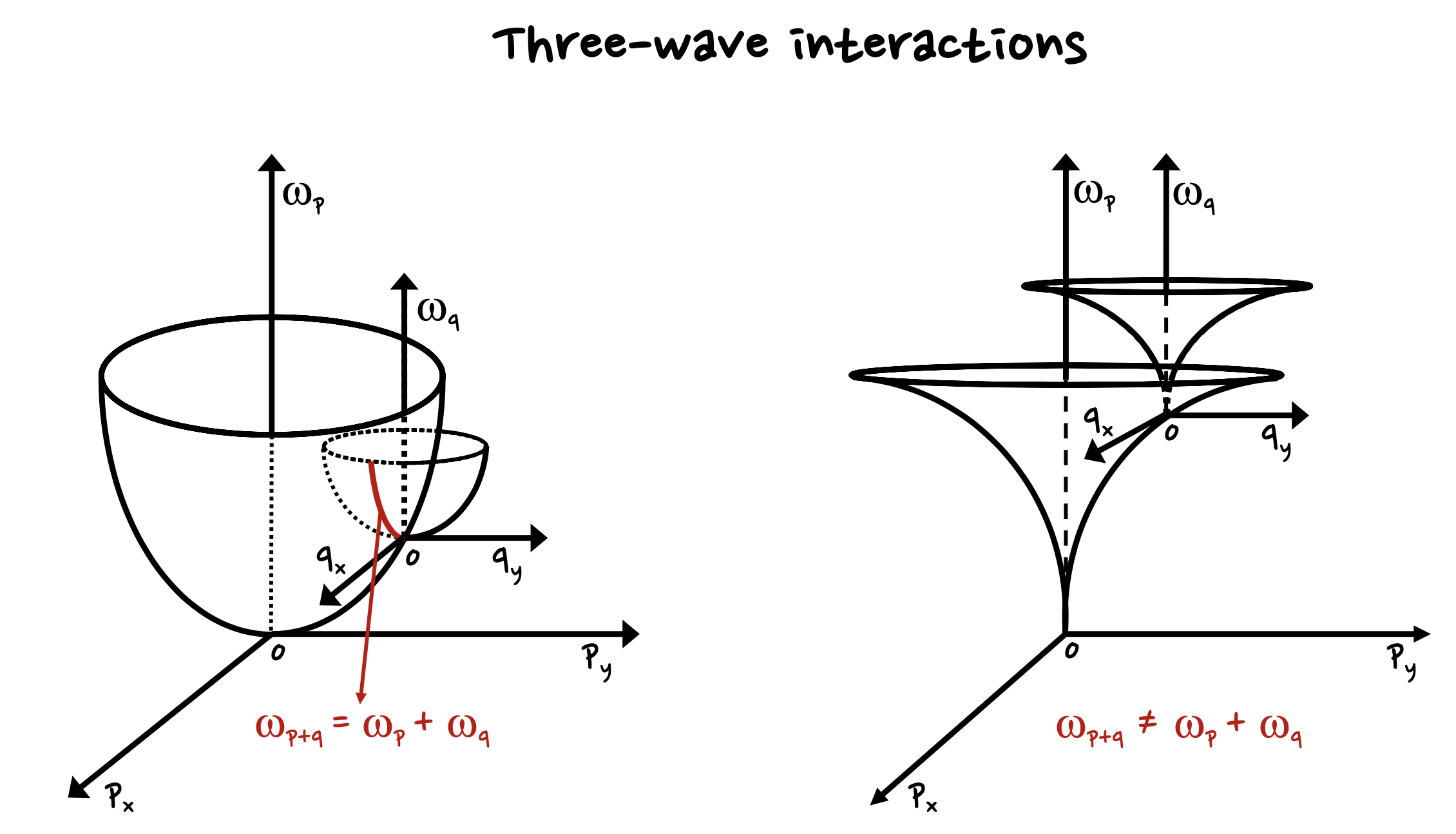}}
\caption{Resonance condition for a dispersion relation of the type $\omega_{k} \sim \vert \kk \vert^{x}$ with $x>1$ (left) and $0<x<1$ (right).}
\label{Fig6-2}
\end{figure}

The triadic resonance condition (\ref{cr2425z}) does not always have a solution. To realize this, we can represent this condition geometrically in the two-dimensional and isotropic case: Figure \ref{Fig6-2} shows a (axisymmetric) dispersion relation of the type $\omega_{k} \sim \vert \kk \vert^{x}$, with $x>1$ (left) and $0<x<1$ (right). The solutions of the resonance condition (\ref{cr2425z}) correspond to the intersection between the surface $\omega(\pp)$ (which is identified with that of $\omega(\kk)$) and the surface $\omega(\qq)$. We find that this intersection exists only for a convex dispersion relation ($x>1$). The concave case corresponding to $0<x<1$ is the one encountered a priori with gravity waves. Note that for anisotropic problems like inertial wave turbulence this simple picture does not work. 

%%%%%%%%%%%%%%%%%%%%%%%%%%%%%%%%%%%%%%%%%%%%%%%%%%%%%%
%%%%%%%%%%%%%%%%%%%%%%%%%%%%%%%%%%%%%%%%%%%%%%%%%%%%%%

The special case $x=1$ is that of non-dispersive (or semi-dispersive) waves. We can easily see that the only possible solution is that where all three wave vectors are aligned. Although a solution to the triadic resonance condition exists, an asymptotic closure of the hierarchy of equations is not guaranteed. Physically, we can easily understand this problem: two semi-dispersive waves propagate at the same speed, therefore, if they initially overlap their interactions will be strong, otherwise they will never interact. Using the multiple time scale method presented in the next section, it can be shown that, in one dimension, a contribution from non-linear terms is possible on a time scale of the order of $T_{1}$ while the closure is made on a longer time scale in $T_{2}$ \cite{Benney1966}. The appearance of these secular contributions requires a new analysis to suppress them and justify the application of the wave turbulence theory. This situation can be found in acoustic wave turbulence or in fast magneto-acoustic wave turbulence in three dimensions  \cite{Galtier2023}. 

Note that for four-wave interactions, the situation is less constraining because new possibilities exist. For example, gravity waves can be analyzed at this order. In the case of non-dispersive waves, it may be possible to develop a theory of wave turbulence because the associated four wave vectors are not necessarily collinear. An example is given by gravitational waves for which the dispersion relation is $\omega_{k}=c k$, with $c$ the speed of light \cite{Galtier2017}.

%%%%%%%%%%%%%%%%%%%%%%%%%%
\subsection{Multiple time scale method}
%%%%%%%%%%%%%%%%%%%%%%%%%%
Following our analysis of the Duffing equation, we wish to make a development of equation (\ref{eq6}) following the method of multiple time scales \cite{Poincare1893}. We introduce a sequence of time scales, $T_{0}$, $T_{1}$, $T_{2}$, ..., which will be treated as independent variables, with
\be
T_{0} \equiv t, \quad T_{1}\equiv \epsilon t, \quad T_{2}\equiv \epsilon^{2} t, ... \, .
\ee
The variation of the wave amplitude with $T_1$ and $T_2$ represents the slow variation that we wish to extract. 
Note that it is because of the weak dependence in $t$ of $T_{1}$, $T_{2}$, ..., that all these variables can be treated (it is an approximation) as independent. 
(The smaller $\epsilon$ is, the better the approximation.) We obtain ($T_0$ being replaced by $t$)
\ba
\left({\partial \over \partial t} + \epsilon {\partial \over \partial T_{1}} + \epsilon^{2} {\partial \over \partial T_{2}} + ...\right) a^{s_k}_{k} =  \epsilon \sum_{s_{p} s_{q}} \int_{\mathbb{R}^{2d}} \HH_{-\kk \pp \qq} a^{s_{p}}_{p} a^{s_{q}}_{q} e^{i \Omega_{k,pq}t} \delta_{k,pq} 
d\pp d\qq \, . && 
\label{eq6bis}
\ea
The variable $a^{s_k}_{k}$ must also be expanded to the power of $\epsilon$ and make the various scales appear in time
\be \label{adevl}
a^{s_k}_{k} = \sum_{n=0}^{+\infty} \epsilon^n a^{s_k}_{k,n}(t,T_{1},T_{2},...) = a^{s_k}_{k,0} + \epsilon a^{s_k}_{k,1} + \epsilon^{2} a^{s_k}_{k,2} + ... \, .
\ee
%The initial condition will be given only by the leading order amplitude $a^{s_k}_{k,0}$, while the other lower contributions will be excited nonlinearly. The Gaussian assumption will be made for the initial condition ($t=0$) to ensure the absence of coherent structures. This is a classical initial condition in DNS when the wave turbulence regime is studied. 
Expression (\ref{adevl}) is then introduced into the fundamental equation (\ref{eq6bis}). We obtain for the first three terms
\begin{subequations}
\begin{align}
{\partial a^{s_k}_{k,0} \over \partial t} &= 0 \, , \\
{\partial a^{s_k}_{k,1} \over \partial t} &= - {\partial a^{s_k}_{k,0} \over \partial T_{1}} + 
\sum_{s_{p} s_{q}} \int_{\mathbb{R}^{2d}} \HH_{-\kk \pp \qq} a_{p,0}^{s_{p}} a_{q,0}^{s_{q}} e^{i \Omega_{k,pq}t} \delta_{k,pq} d\pp d\qq \, , \\
{\partial a^{s_k}_{k,2} \over \partial t} &= - {\partial a^{s_k}_{k,1} \over \partial T_{1}} - {\partial a^{s_k}_{k,0} \over \partial T_{2}} + 
\sum_{s_{p} s_{q}} \int_{\mathbb{R}^{2d}} \HH_{-\kk \pp \qq} \left[ a_{p,1}^{s_{p}} a_{q,0}^{s_{q}} + a_{p,0}^{s_{p}} a_{q,1}^{s_{q}} \right] e^{i \Omega_{k,pq}t} \delta_{k,pq} d\pp d\qq \, . 
\end{align}
\end{subequations}
To lighten the writing, the time dependency of the variables has been omitted. After integration on $t$, one finds
\begin{subequations}
\begin{align}
a^{s_k}_{k,0} &= a^{s_k}_{k,0}(T_1,T_2,...) \, , \label{137} \\
a^{s_k}_{k,1} &= -t {\partial a^{s_k}_{k,0} \over \partial T_{1}} + b^{s_k}_{k,1} \, , \label{138} \\
a^{s_k}_{k,2} &= \frac{t^2}{2} {\partial^2 a^{s_k}_{k,0} \over \partial T_{1}^2} -t {\partial a^{s_k}_{k,0} \over \partial T_{2}} 
- \int_0^t {\partial b^{s_k}_{k,1} \over \partial T_{1}} dt + \tilde b^{s_k}_{k,2} \, , \label{139} 
\end{align}
\end{subequations}
with by definition
\be \label{bdelta}
b^{s_k}_{k,1} \equiv \sum_{s_{p} s_{q}} \int_{\mathbb{R}^{2d}} \HH_{-\kk \pp \qq} a_{p,0}^{s_{p}} a_{q,0}^{s_{q}} \Delta(\Omega_{k,pq}) \delta_{k,pq} d\pp d\qq \, , 
\ee
\ba
\Delta(X) \equiv \int_{0}^{t} e^{i Xt} dt  = \frac{e^{i Xt} -1}{i X} ,
\ea
and
\be \label{bdelta2a}
\tilde b^{s_k}_{k,2} \equiv \sum_{s_{p} s_{q}} \int_{\mathbb{R}^{2d}} \HH_{-\kk \pp \qq} \int_{0}^{t} 
\left[ a_{p,1}^{s_{p}} a_{q,0}^{s_{q}} + a_{p,0}^{s_{p}} a_{q,1}^{s_{q}}\right] e^{i \Omega_{k,pq}t} dt \delta_{k,pq} d\pp d\qq \, .
\ee
The previous equation is modified when expression (\ref{138}) is introduced; one finds
$$
\tilde b^{s_k}_{k,2} = -\sum_{s_{p} s_{q}} \int_{\mathbb{R}^{2d}} \HH_{-\kk \pp \qq} 
\frac{\partial (a^{s_p}_{p,0} a_{q,0}^{s_{q}})}{\partial T_{1}} \left( \int_0^t t e^{i \Omega_{k,pq}t} dt \right) \delta_{k,pq} d\pp d\qq
$$
\be
+ \sum_{s_{p} s_{q}} \int_{\mathbb{R}^{2d}} \HH_{-\kk \pp \qq} 
\left( \int_{0}^{t} (b^{s_p}_{p,1} a_{q,0}^{s_{q}} + b^{s_q}_{q,1} a_{p,0}^{s_{p}} ) e^{i \Omega_{k,pq}t} dt \right) 
\delta_{k,pq} d\pp d\qq \, .
\ee
Expression (\ref{139}) becomes
\be \label{34e}
a^{s_k}_{k,2} = \frac{t^2}{2} {\partial^2 a^{s_k}_{k,0} \over \partial T_{1}^2} -t {\partial a^{s_k}_{k,0} \over \partial T_{2}} + b^{s_k}_{k,2} \, , 
\ee
with
$$
b^{s_k}_{k,2} = -\sum_{s_{p} s_{q}} \int_{\mathbb{R}^{2d}} \HH_{-\kk \pp \qq} \frac{\partial (a^{s_p}_{p,0} a_{q,0}^{s_{q}})}{\partial T_{1}} 
\left( \int_0^t \left[\Delta( \Omega_{k,pq}) + t e^{i \Omega_{k,pq}t} \right] dt \right)  \delta_{k,pq} d\pp d\qq
$$
$$
+ 2 \sum_{s_{p} s_{q} s_{p'} s_{q'} } \int_{\mathbb{R}^{4d}} \HH_{-\kk \pp \qq} \HH_{-\pp \pp' \qq'} a_{p',0}^{s_{p'}} a_{q',0}^{s_{q'}}  a_{q,0}^{s_{q}} 
\left( \int_{0}^{t} \Delta( \Omega_{p,p'q'}) e^{i \Omega_{k,pq}t} dt \right) \delta_{k,pq} \delta_{p,p'q'} d\pp d\qq d\pp' d\qq' \, .
$$
The time integrals give the relations
\ba
\int_0^t \left[\Delta( \Omega_{k,pq}) + t e^{i \Omega_{k,pq}t} \right] dt &=& t \Delta( \Omega_{k,pq}) \, , \\
\int_{0}^{t} \Delta( \Omega_{p,p'q'}) e^{i \Omega_{k,pq}t} dt &=& \frac{\Delta(\Omega_{k,p'q'q}) - \Delta(\Omega_{k,pq})}{i (\Omega_{k,p'q'q}-\Omega_{k,pq})} \, ,
\ea
that will be used below in the long time limit.

%%%%%%%%%%%%%%%%%%%%%%%%%%
\subsection{First asymptotic closure at time $T_1$}
%%%%%%%%%%%%%%%%%%%%%%%%%%

With the previous definitions, the perturbative expansion of the second-order moment writes
\ba \label{37ez}
\langle a^{s_k}_{k} a^{s_{k'}}_{k'} \rangle &=& 
\langle (a^{s_k}_{k,0} + \epsilon a^{s_k}_{k,1} + \epsilon^2 a^{s_k}_{k,2} + ...)(a^{s_{k'}}_{k',0} + \epsilon a^{s_{k'}}_{k',1} + \epsilon^2 a^{s_{k'}}_{k',2} + ...) \rangle \\
&=& \langle a^{s_k}_{k,0} a^{s_{k'}}_{k',0} \rangle + \epsilon \langle a^{s_k}_{k,0} a^{s_{k'}}_{k',1} + a^{s_k}_{k,1} a^{s_{k'}}_{k',0} \rangle 
+ \epsilon^2 \langle a^{s_k}_{k,0} a^{s_{k'}}_{k',2} + a^{s_k}_{k,1} a^{s_{k'}}_{k',1} + a^{s_k}_{k,2} a^{s_{k'}}_{k',0}  \rangle + ... \nonumber 
\ea
We shall assume that this turbulence is statistically homogeneous. In this case, the second-order moment can be written in term of second-order cumulant $q^{s_ks_{k'}}(\kk,\kk') \equiv q_{k}^{s_ks_{k'}}$ such that 
\be
\langle a^{s_k}_{k} a^{s_{k'}}_{k'} \rangle = q_{k}^{s_ks_{k'}} \delta (\kk+\kk') \, ,
\ee
where the presence of $\delta (\kk+\kk')$ is the consequence of the statistical homogeneity \cite{GaltierCUP2023}. 
We also assume -- this is the basic idea of the method -- that the second-order moment (in fact, the coefficient $q_{k}^{s_ks_{k'}}$ in front of the Dirac function) in the left hand side remains bounded at all time \cite{Benney1966}. 
As an example, we can think of the energy spectrum which, as we know, remains physically bounded. Therefore, the contributions on the right hand side must also be bounded (in the sense introduced above). We will see that secular terms can appear at different orders in $\epsilon$; this leads to certain conditions to cancel their contributions in order to keep the development uniform in time. As we shall see, at order $\mathcal{O} (\epsilon^2)$ this condition leads to the so-called kinetic equations. 

At order $\mathcal{O} (\epsilon^0)$, we have the contribution of $\langle a^{s_k}_{k,0} a^{s_{k'}}_{k',0} \rangle$ which will therefore be assumed to be bounded at all times. 

At order $\mathcal{O} (\epsilon^1)$, we have the contribution
\ba \label{38e}
\langle a^{s_k}_{k,0} a^{s_{k'}}_{k',1} + a^{s_k}_{k,1} a^{s_{k'}}_{k',0} \rangle &=& 
\left\langle a^{s_k}_{k,0} \left(-t {\partial a^{s_{k'}}_{k',0} \over \partial T_{1}} + b^{s_{k'}}_{k',1}\right) + \left(-t {\partial a^{s_k}_{k,0} \over \partial T_{1}} + b^{s_k}_{k,1}\right) a^{s_{k'}}_{k',0} \right\rangle \nonumber \\
&=& -t \frac{\partial}{\partial T_{1}} \langle a^{s_k}_{k,0} a^{s_{k'}}_{k',0} \rangle + \langle a^{s_k}_{k,0} b^{s_{k'}}_{k',1} 
+ b^{s_k}_{k,1} a^{s_{k'}}_{k',0} \rangle \, .
\ea
The first term on the right hand side gives a secular contribution proportional to $t$. For the second term, we have
\be \label{39e}
\sum_{s_{p} s_{q}} \int_{\mathbb{R}^{2d}} \HH_{-\kk \pp \qq} \langle a^{s_k}_{k,0} a_{p,0}^{s_{p}} a_{q,0}^{s_{q}} \rangle \Delta(\Omega_{k,pq}) \delta_{k,pq} d\pp d\qq \, .
\ee
The long time limit ($t \gg 1/\omega$) of this oscillating integral will be given by the Riemann-Lebesgue lemma (the proof requires to use of generalized functions)
\ba
\Delta(X) &=& \frac{e^{i Xt} -1}{i X} \xrightarrow{\text{t $\to +\infty$}} \pi \delta(X) + i {\cal P} \left(\frac{1}{X}\right) \, , \label{RiemannL}
\ea
where ${\cal P}$ is the Cauchy principal value of the integral. Therefore, the long time limit of expression (\ref{39e}) gives no secular contribution. The same conclusion is obtained for the third term of expression (\ref{38e}). Also, the condition to cancel the unique secular term is
\be
\frac{\partial \langle a^{s_k}_{k,0} a^{s_{k'}}_{k',0} \rangle }{\partial T_{1}} = 0 \, ,
\ee
which means that the second-order moment does not evolve over a time scale $T_1$. As will be seen later, a turbulent cascade is only expected on a time scale $T_2$.

%%%%%%%%%%%%%%%%%%%%%%%%%%
\subsection{Second asymptotic closure at time $T_2$}
%%%%%%%%%%%%%%%%%%%%%%%%%%

The analysis continues at order $\mathcal{O} (\epsilon^2)$. With expression (\ref{34e}), the next contribution reads
\ba
&&\left\langle a^{s_k}_{k,1} a^{s_{k'}}_{k',1} + a^{s_k}_{k,0} a^{s_{k'}}_{k',2} + a^{s}_{k,2} a^{s_{k'}}_{k',0} \right\rangle = 
\left\langle \left(-t {\partial a^{s_k}_{k,0} \over \partial T_{1}} + b^{s_k}_{k,1}\right)\left(-t {\partial a^{s_{k'}}_{k',0} \over \partial T_{1}} + b^{s_{k'}}_{k',1}\right) \right\rangle \\
&& \quad + \left\langle a^{s_k}_{k,0} \left(\frac{t^2}{2} {\partial^2 a^{s_{k'}}_{k',0} \over \partial T_{1}^2} -t {\partial a^{s_{k'}}_{k',0} \over \partial T_{2}} 
+ b^{s_{k'}}_{k',2} \right) + a^{s_{k'}}_{k',0} \left(\frac{t^2}{2} {\partial^2 a^{s_k}_{k,0} \over \partial T_{1}^2} -t {\partial a^{s_k}_{k,0} \over \partial T_{2}} + b^{s_k}_{k,2} \right) \right\rangle \, , \nonumber 
\ea
which gives after development and simplifications
\ba \label{eq44e}
\left\langle a^{s_k}_{k,1} a^{s_{k'}}_{k',1} + a^{s_k}_{k,0} a^{s_{k'}}_{k',2} + a^{s_k}_{k,2} a^{s_{k'}}_{k',0} \right\rangle &=& 
\frac{t^2}{2} \frac{\partial^2 \langle a^{s_k}_{k,0} a^{s_{k'}}_{k',0} \rangle }{\partial T_{1}^2} 
- t \frac{\partial \langle a^{s_k}_{k,0} a^{s_{k'}}_{k',0} \rangle }{\partial T_{2}} 
+ \left\langle b^{s_k}_{k,1} b^{s_{k'}}_{k',1} \right\rangle \\
&-& t \left\langle {\partial a^{s_k}_{k,0} \over \partial T_{1}} b^{s_{k'}}_{k',1} + b^{s_k}_{k,1} {\partial a^{s_{k'}}_{k',0} \over \partial T_{1}} \right\rangle 
+ \left\langle a^{s_k}_{k,0} b^{s_{k'}}_{k',2} + a^{s_{k'}}_{k',0} b^{s_k}_{k,2} \right\rangle \, . \nonumber 
\ea
The first term on the right hand side cancels over the long time as required by the first asymptotic closure. The second term gives a secular contribution. The other three terms can potentially give a secular contribution: it is obvious for the fourth term and non-trivial for the third and fifth terms which require further development. 

The third term on the right hand side writes
\ba \label{46e}
\left\langle b^{s_k}_{k,1} b^{s_{k'}}_{k',1} \right\rangle &=& 
\sum_{s_{p} s_{q} s_{p'} s_{q'}} \int_{\mathbb{R}^{4d}} \HH_{-\kk \pp \qq} \HH_{-\kk' \pp' \qq'} 
\langle a_{p,0}^{s_{p}} a_{q,0}^{s_{q}} a_{p',0}^{s_{p'}} a_{q',0}^{s_{q'}} \rangle \Delta(\Omega_{k,pq}) \Delta(\Omega_{k',p'q'}) \nonumber \\
&&\delta_{k,pq} \delta_{k',p'q'} d\pp d\qq d\pp' d\qq' \, , 
\ea
Here again, the theory of generalized functions gives us the long time behavior of this oscillating integral (with the Poincar\'e-Bertrand formula) 
\ba
\Delta(X) \Delta(-X) \xrightarrow{\text{t $\to +\infty$}} 2 \pi t \delta(X) + 2 {\cal P} \left(\frac{1}{X}\right) \frac{\partial}{\partial X} \, .
\ea
Therefore, a secular contribution involving a Dirac function is possible. The fourth-order moment in expression (\ref{46e}) can be decomposed into products of second-order cumulant plus a fourth-order cumulant such that (the statistical homogeneity is used as well as $\langle a_{k,0}^{s_k}\rangle = 0$)
\ba  \label{stat33}
\langle a_{p,0}^{s_{p}} a_{q,0}^{s_{q}} a_{p',0}^{s_{p'}} a_{q',0}^{s_{q'}} \rangle &=&  
q^{s_{p}s_{q}s_{p'}s_{q'}}_{pqp',0} \delta(\pp+\qq+\pp'+\qq') 
+ q^{s_{p} s_q}_{p,0} q^{s_{p'} s_{q'}}_{p',0} \delta(\pp+\qq) \delta(\pp'+\qq')  \\
&+& q^{s_{p} s_{p'}}_{p,0} q^{s_{q} s_{q'}}_{q,0} \delta(\pp+\pp') \delta(\qq+\qq') 
+ q^{s_{p} s_{q'}}_{p,0} q^{s_{q} s_{p'}}_{q,0} \delta(\pp+\qq') \delta(\qq+\pp') \, . \nonumber
\ea
Note that according to expression (\ref{37ez}) by homogeneity we also have the relation $\kk=-\kk'$. We obtain 
\ba \label{48e}
&&\left\langle b^{s_k}_{k,1} b^{s_{k'}}_{k',1} \right\rangle = 
\sum_{s_{p} s_{q} s_{p'} s_{q'}} \int_{\mathbb{R}^{4d}} \HH_{-\kk \pp \qq} \HH_{-\kk' \pp' \qq'} 
\left[ q^{s_{p}s_{q}s_{p'}s_{q'}}_{pqp',0} \delta(\pp+\qq+\pp'+\qq') \right.  \\
&& + q^{s_{p} s_q}_{p,0} q^{s_{p'} s_{q'}}_{p',0} \delta(\pp+\qq) \delta(\pp'+\qq') 
+ q^{s_{p} s_{p'}}_{p,0} q^{s_{q} s_{q'}}_{q,0} \delta(\pp+\pp') \delta(\qq+\qq') \nonumber \\
&&\left. + q^{s_{p} s_{q'}}_{p,0} q^{s_{q} s_{p'}}_{q,0} \delta(\pp+\qq') \delta(\qq+\pp') \right] 
\Delta(\Omega_{k,pq}) \Delta(\Omega_{k',p'q'}) \delta_{k,pq} \delta_{k',p'q'} d\pp d\qq d\pp' d\qq' \, . \nonumber
\ea
We are looking for secular contributions. In the second line, the first term does not contribute since it imposes $\kk={\bf 0}$ which cancels $\HH_{\kk \pp \qq}$ but the second term can contribute when the conditions $s_p=-s_{p'}$ and $s_q=-s_{q'}$ are satisfied. Likewise, in the third line a contribution is possible when $s_p=-s_{q'}$ and $s_q=-s_{p'}$. There is no contribution from the first line, which means that the situation is the same as if the distribution were Gaussian (however, we do not make this assumption). Finally, in the long time limit, the secular contribution ${\cal C}_t \left\langle b^{s_k}_{k,1} b^{s_{k'}}_{k',1} \right\rangle$ is 
\ba \label{49e}
{\cal C}_t  \left\langle b^{s_k}_{k,1} b^{s_{k'}}_{k',1} \right\rangle &=& 
4 \pi t \sum_{s_{p} s_{q}} \int_{\mathbb{R}^{2d}} \HH_{-\kk \pp \qq} \HH_{\kk -\pp -\qq} 
q^{s_{p} -s_{p}}_{p,0} q^{s_{q} -s_{q}}_{q,0} \delta (\Omega_{k,pq}) \delta_{k,pq} \delta_{kk'} d\pp d\qq \nonumber\\
&=& 4 \pi t \sum_{s_{p} s_{q}} \int_{\mathbb{R}^{2d}} \vert \HH_{-\kk \pp \qq} \vert^2
q^{s_{p} -s_{p}}_{p,0} q^{s_{q} -s_{q}}_{q,0} \delta (\Omega_{k,pq}) \delta_{k,pq} \delta_{kk'} d\pp d\qq \, .
\ea

The fourth term on the right hand side of equation (\ref{eq44e}) does not contribute over the long time because it depends on the $T_1$ derivative. 
The proof is given by a new relation involving the n-order moments which can be written
\ba
\langle a_k^{s_k} a_{k'}^{s_{k'}} a_{k''}^{s_{k''}} ... \rangle &=& \langle a_{k,0}^{s_k} a_{k',0}^{s_{k'}} a_{k'',0}^{s_{k''}} ... \rangle \\
&+& \epsilon \langle a_{k,1}^{s_k} a_{k',0}^{s_{k'}} a_{k'',0}^{s_{k''}} ... + a_{k,0}^{s_k} a_{k',1}^{s_{k'}} a_{k'',0}^{s_{k''}} ... 
+ a_{k,0}^{s_k} a_{k',0}^{s_{k'}} a_{k'',1}^{s_{k''}} .... + ....\rangle \nonumber \\
&+& \epsilon^2 \langle .... \rangle + ... \nonumber
\ea
As for the second-order moment, we demand that the moments of order n are bounded. At order $\mathcal{O} (\epsilon)$, we obtain the relation
\ba
&&\langle a_{k,1}^{s_k} a_{k',0}^{s_{k'}} a_{k'',0}^{s_{k''}} ... + a_{k,0}^{s_k} a_{k',1}^{s_{k'}} a_{k'',0}^{s_{k''}} ... 
+ a_{k,0}^{s_k} a_{k',0}^{s_{k'}} a_{k'',1}^{s_{k''}} ... + ...\rangle \\
&& = \left\langle \left(-t {\partial a^{s_k}_{k,0} \over \partial T_{1}} + b^{s_k}_{k,1}\right) a_{k',0}^{s_{k'}} a_{k'',0}^{s_{k''}} ... 
+ a_{k,0}^{s_k} \left(-t {\partial a^{s_{k'}}_{k',0} \over \partial T_{1}} + b^{s_{k'}}_{k',1}\right) a_{k'',0}^{s_{k''}} ... + ... \right\rangle \nonumber \\
&& = -t {\partial \langle a^{s_k}_{k,0} a^{s_{k'}}_{k',0} a_{k'',0}^{s_{k''}} ... \rangle \over \partial T_{1}} 
+ \langle b^{s_k}_{k,1}a_{k',0}^{s_{k'}} a_{k'',0}^{s_{k''}} ... \rangle + \langle a^{s_k}_{k,0} b^{s_{k'}}_{k',1} a_{k'',0}^{s_{k''}} ... \rangle + ... \nonumber
\ea
We see that only the first term of the last line gives a secular contribution over the long time, which means that we have to satisfy the condition
\be
{\partial \langle a^{s_k}_{k,0} a^{s_{k'}}_{k',0} a_{k'',0}^{s_{k''}} ... \rangle \over \partial T_{1}} = 0 \, ,
\ee
at any order n. Therefore, the probability density function does not depend on $T_1$ and we can assume that the variable itself does not depend on $T_1$.  
It is a mild hypothesis because it is difficult to imagine a turbulent system where everything fluctuates, and in which it would be possible to have a $T_1$ dependence for the amplitude whereas the probability density function has no such dependence. 
This shows that the fourth term on the right hand side of equation (\ref{eq44e}) does not contribute in the long time limit. 

The last term of equation (\ref{eq44e}) writes 
\be
\left\langle a^{s_k}_{k,0} b^{s_{k'}}_{k',2} + a^{s_{k'}}_{k',0} b^{s_k}_{k,2} \right\rangle = 
\ee
$$
\sum_{s_{p} s_{q} s_{p'} s_{q'} } \int_{\mathbb{R}^{4d}} 2 \HH_{-\kk' \pp \qq} \HH_{-\pp \pp' \qq'} 
\left\langle a^{s_k}_{k,0} a_{p',0}^{s_{p'}} a_{q',0}^{s_{q'}}  a_{q,0}^{s_{q}} \right\rangle
\left( \frac{\Delta(\Omega_{k',p'q'q}) - \Delta(\Omega_{k',pq})}{i (\Omega_{k',p'q'q}-\Omega_{k',pq})} \right) 
\delta_{k',pq} \delta_{p,p'q'} d\pp d\qq d\pp' d\qq'
$$
$$
+\sum_{s_{p} s_{q} s_{p'} s_{q'} } \int_{\mathbb{R}^{4d}} 2 \HH_{-\kk \pp \qq} \HH_{-\pp \pp' \qq'} 
\left\langle a^{s_{k'}}_{k',0} a_{p',0}^{s_{p'}} a_{q',0}^{s_{q'}}  a_{q,0}^{s_{q}} \right\rangle
\left( \frac{\Delta(\Omega_{k,p'q'q}) - \Delta(\Omega_{k,pq})}{i (\Omega_{k,p'q'q}-\Omega_{k,pq})} \right) 
\delta_{k,pq} \delta_{p,p'q'} d\pp d\qq d\pp' d\qq' .
$$
The secular contributions of these oscillating integrals will be given by the theory of generalized functions with the relation
\be
\frac{\Delta(X) - \Delta(0)}{iX} \xrightarrow{\text{t $\to +\infty$}} \pi t \delta(X) + it {\cal P} \left(\frac{1}{X}\right) \, .
\ee
We also need to use the following development (and its symmetric in $\kk'$)
\ba  \label{55e}
\langle a_{k,0}^{s_k} a_{p',0}^{s_{p'}} a_{q',0}^{s_{q'}} a_{q,0}^{s_{q}} \rangle &=&  
q^{s_{k}s_{p'}s_{q'}s_{q}}_{kp'q',0} \delta(\kk+\pp'+\qq'+\qq) 
+ q^{s_k s_{q}}_{k,0} q^{s_{p'} s_{q'}}_{p',0} \delta(\kk+\qq) \delta(\pp'+\qq') \\
&+& q^{s_k s_{q'}}_{k,0} q^{s_{p'} s_{q}}_{p',0} \delta(\kk+\qq') \delta(\pp'+\qq) 
+ q^{s_k s_{p'}}_{k,0} q^{s_{q'} s_{q}}_{q',0} \delta(\kk+\pp') \delta(\qq'+\qq)  \, . \nonumber
\ea
In the right hand side of expression (\ref{55e}), the first two terms do not give a secular contribution, however, the last two terms give a contribution when the following conditions are satisfied, namely $s_k=-s_{q'}$, $s_{p'}=-s_q$ and $s_k=-s_{p'}$, $s_{q'}=-s_q$, respectively. After substitution and simplification, we obtain the secular contributions in the long time limit
\ba 
&&{\cal C}_t \left\langle a^{s_k}_{k,0} b^{s_{k'}}_{k',2} + a^{s_{k'}}_{k',0} b^{s_k}_{k,2} \right\rangle = \\
&&+4t \sum_{s_{p} s_{q}} \int_{\mathbb{R}^{2d}} \HH_{-\kk' \pp \qq} \HH_{-\pp -\qq \kk'} 
q^{s_k s_{k'}}_{k,0} q^{s_{q} -s_{q}}_{q,0}
\left( \pi \delta(\Omega_{k',pq}) + i {\cal P} \left(\frac{1}{\Omega_{k',pq}}\right) \right) 
\delta_{k',pq} \delta_{kk'} d\pp d\qq \nonumber \\
&& +4t \sum_{s_{p} s_{q}} \int_{\mathbb{R}^{2d}} \HH_{-\kk \pp \qq} \HH_{-\pp -\qq \kk} 
q^{s_k s_{k'}}_{k,0} q^{s_{q} -s_{q}}_{q,0}
\left( \pi \delta(\Omega_{k,pq}) + i {\cal P} \left(\frac{1}{\Omega_{k,pq}}\right) \right) 
\delta_{k,pq} \delta_{kk'} d\pp d\qq \nonumber \\
&&= 8 \pi t \sum_{s_{p} s_{q}} \int_{\mathbb{R}^{2d}} \HH_{-\kk \pp \qq} \HH_{-\pp -\qq \kk} 
q^{s_k s_{k'}}_{k,0} q^{s_{q} -s_{q}}_{q,0}
\delta(\Omega_{k,pq}) \delta_{k,pq} \delta_{kk'} d\pp d\qq \nonumber .
\ea
The last writing is obtained by using the property $\HH^*_{-\kk \pp \qq} = \HH_{-\kk \pp \qq}$ and the symmetry in $\pp \to -\pp$ and $\qq \to -\qq$. 

If we impose the nullity of the sum of the different secular contributions at time $T_2$, we find the condition
\ba
\frac{\partial \langle a^{s_k}_{k,0} a^{s_{k'}}_{k',0} \rangle }{\partial T_{2}} &=&
\frac{\partial q^{s_ks_{k'}}_{k,0}}{\partial T_{2}} \delta_{kk'} \\
&=& 4 \pi \sum_{s_{p} s_{q}} \int_{\mathbb{R}^{2d}} \vert \HH_{-\kk \pp \qq} \vert^2
q^{s_{p} -s_{p}}_{p,0} q^{s_{q} -s_{q}}_{q,0} \delta (\Omega_{k,pq}) \delta_{k,pq} \delta_{kk'} d\pp d\qq \nonumber \\
&+& 8 \pi \sum_{s_{p} s_{q}} \int_{\mathbb{R}^{2d}} \HH_{-\kk \pp \qq} \HH_{-\pp -\qq \kk} 
q^{s_k s_{k'}}_{k,0} q^{s_{q} -s_{q}}_{q,0} \delta(\Omega_{k,pq}) \delta_{k,pq} \delta_{kk'} d\pp d\qq \, . \nonumber
\ea
After integrating over $\kk'$, using the property that $q^{s_ks_{k'}}_{k,0}$ is real and thus $s_k=-s_{k'}$ and introducing $N^{s_k}_k \equiv q^{s_k -s_{k}}_{k,0}$, we end up with (some simple manipulations are also used to symmetrize the equation)
\be \label{ke58}
\frac{\partial N_k^{s_k}}{\partial t} = 
\ee
$$
4 \pi \epsilon^2 \sum_{s_{p} s_{q}} \int_{\mathbb{R}^{2d}} \HH_{\kk \pp \qq}
\left[ \frac{\HH_{-\kk -\pp -\qq}}{N_k^{s_k}} + \frac{\HH_{-\pp -\qq -\kk}}{N_p^{s_p}} + \frac{\HH_{-\qq -\pp -\kk}}{N_q^{s_q}} \right]
N_k^{s_k} N_p^{s_p} N_q^{s_q} \delta (\Omega_{kpq}) \delta_{kpq} d\pp d\qq \, , 
$$
which is the kinetic equation for three-wave interactions.

%%%%%%%%%%%%%%%%%%%%%%%%%%
\subsection{Conclusion}
%%%%%%%%%%%%%%%%%%%%%%%%%%

The derivation of the kinetic equation (\ref{ke58}) does not involve any closure assumption such as the quasi-Gaussianity and the random phase appears naturally dynamically in wave turbulence. The kinetic equation is obtained by using a sequential asymptotic closures at times $T_1$ and $T_2$. This method consists in imposing the nullity of secular terms (proportional to $t$) which emerge over long times in order to guarantee a bounded value for the associated moments. These secular terms do not involve fourth-order cumulants but only products of second-order cumulants. 
Note, however, that the derivation made, although systematic, says nothing about the remaining terms (not used to obtain the kinetic equation) in the small $\epsilon$ perturbation expansion, the implicit conjecture being that they are subdominant. The proof of this conjecture remains a mathematical challenge \cite{Deng2021}. 

In summary, it can be said that wave turbulence for three-wave interactions is characterized by a dynamics on two time scales. On short timescales, of the order of the wave period, there is a phase mixing which leads, due to the dispersive nature of the waves, to the decoupling of the correlations if initially present and to a statistics close to Gaussianity, as expected from the central limit theorem. On a longer time scale, the non-linear coupling -- weak at short times -- becomes non-negligible because of the resonance mechanism. This coupling leads to a regeneration of the cumulants via the product of lower order cumulants. It is these terms that are at the origin of the energy transfer mechanism. 

In this derivation, the system studied is assumed to be of infinite size and can therefore be treated as continuous. Note that numerical simulation with its grid of points escapes this description. Effects (freezing of the cascade) linked to the discretization of the Fourier space can appear because the resonance conditions are a priori more difficult to satisfy (see for example \cite{Connaughton2001} for capillary waves). In theory, the weaker the non-linearities, the stronger these effects are. For example, in inertial wave turbulence \cite{Bouroudia2008} has shown that discretization effects become preponderant when the Rossby number, $R_{o}$, is smaller than $10^{-3}$. Beyond this value, but still for a small $R_{o} \ll 1$, these effects are negligible because of the quasi-resonances which, together with the resonances, contribute to the transfer of energy. 

In nature we often encounter three-wave problems. As mentioned above, sometimes higher-level problems (up to six waves) can occur. Despite the diversity of situations, the kinetic equation takes a universal form in the sense that it is written (at leading order)
\be
\frac{\partial N^{s_k}_k}{\partial t} = \epsilon^{2n-4} T(\kk) \, ,
\ee
for $n$-wave interactions, with $T$ the transfer function (or collision integral). The higher the degree of interaction, the longer the transfer time  $\tau_{tr}$. This time is characterized by the relation $\omega \tau_{tr} \sim \mathcal{O} (1/\epsilon^{2n-4})$. Since the characteristic time of the waves is $\tau_{\omega} \sim 1/\omega$ and the small parameter is none other than the ratio between this time and the non-linear time $\tau_{NL}$, we arrive at the following phenomenological expression for the transfer time
\be
\tau_{tr} \sim \frac{\tau_{\omega}}{\epsilon^{2n-4}} \sim \frac{\tau_{NL}^{2n-4}}{\tau_{\omega}^{2n-5}} \, .
\ee
Therefore, for three-wave interactions, we can use the expression: $\tau_{tr} \sim \omega \tau^{2}_{NL}$.

%%%%%%%%%%%%%%%%%%
\section{Applications to Navier-Stokes equations}
%%%%%%%%%%%%%%%%%%
\subsection{Capillary wave turbulence}
Together with gravity waves, capillary waves are the main surface waves encountered in nature. The latter have an advantage over the former in that they are easier to treat analytically in the non-linear regime since they involves three-wave interactions. Following the multiple time scale method, we obtain the following kinetic equation for capillary wave turbulence \cite{Zakharov1967,Galtier2020}
\be \label{KEcw}
\frac{\partial N_k^{s_k}}{\partial t} = 
4 \pi \epsilon^{2} \sum_{s_{p} s_{q}} \int_{\RR^{4}} \vert \HH_{kpq} \vert^2 s_{p}s_{q}  
\left[ \frac{s_{p}s_q}{N_k^{s_k}} + \frac{s_k s_q}{N_p^{s_p}} + \frac{s_k s_p}{N_q^{s_q}} \right] 
N_k^{s_k} N_p^{s_p} N_q^{s_q} \delta(\Omega_{kpq}) \delta_{kpq} d\pp d\qq \, ,  
\ee
with
\ba
\vert \HH_{kpq} \vert^{2} = \frac{\sqrt{\sigma}}{8} \left[s_k (\pp \cdot \qq + pq) \left(\frac{pq}{k}\right)^{1/4} 
+ s_{p} (\kk \cdot \qq + kq) \left(\frac{qk}{p}\right)^{1/4} + s_{q} (\kk \cdot \pp + kp) \left(\frac{pk}{q}\right)^{1/4} \right]^{2} ,
\ea
where $\omega_{k} = \sqrt{\sigma} k^{3/2}$ and $\sigma=\gamma / \rho_{water}$, with $\gamma$ the surface tension coefficient (for the air-water interface $\gamma \simeq 0.07$\,N/m) and $\rho_{water}$ the mass density of the water.

\subsection{Inertial wave turbulence}
Inertial wave turbulence is the regime reached by an incompressible fluid subjected to a rapid and uniform rotation within the limit of a large Reynolds number. 
At first sight, this situation seems similar to eddy turbulence since it simply involves adding a well-known force to the Navier-Stokes equations, namely the Coriolis force. However, there are two major differences: waves exist and the spherical symmetry is broken by the presence of a privileged axis, that of rotation. This anisotropy makes the study of inertial wave turbulence more complex than that of capillary waves. Nevertheless, the multiple time scale method can be applied and the following anisotropic kinetic equation can be obtained in the limit $k_\perp \gg k_\parallel$ \cite{Galtier2003}
\ba \label{KEIW}
\frac{\partial N_k^{s}}{\partial t} &=& \frac{\pi \epsilon^2}{4 \, s\omega_k} \sum_{s_{p} s_{q}} \int  
\left( \frac{\sin \theta_k}{\kpn} \right)^2 (s_p \ppn - s_q \qpn)^2 ( s \kpn + s_p \ppn + s_q \qpn)^2 \\
&&\mbox{} \times \left[ s\omega_k N_p^{s_p} N_q^{s_q} + s_p \omega_p N_k^{s} N_q^{s_q} + s_q \omega_q N_k^{s} N_p^{s_p} \right] 
\delta (\Omega_{kpq}) \delta_{kpq} d\pp d\qq ,  \nonumber
\ea
with $\omega_{k} = 2 \Omega_0 k_\parallel / k_\perp$, ${\bf e_\parallel}$ ($\vert {\bf e_\parallel} \vert = 1$) the axis of rotation, $\Omega_0$ the constant rotating rate and $\theta_k$ the opposite angle to $\kk_\perp$ in the triangle $\kk_\perp = \pp_\perp + \qq_\perp$.

%%%%%%%%%%%%%%%%%%
\section{Kolmogorov-Zakharov spectrum}
\subsection{A toy model}
%%%%%%%%%%%%%%%%%%

\begin{figure}[t]
\center
\centerline{\includegraphics[width=1\linewidth]{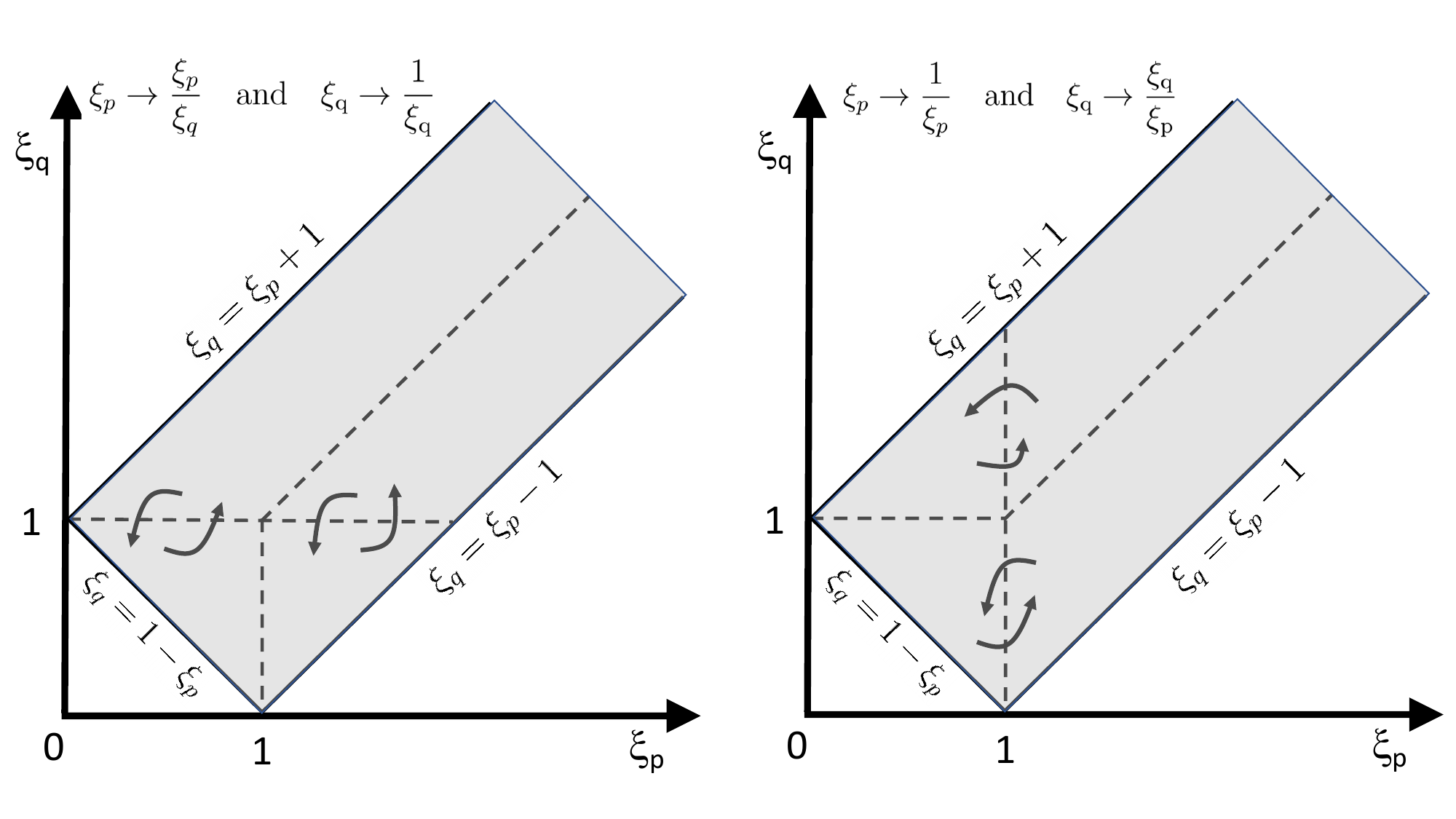}}
\caption{The isotropic kinetic equation (\ref{KEtoy}) involves triadic interactions limited to the infinitely long gray band. The Zakharov transformations (TZ1 and TZ2 are in the right and left panels, respectively) consists in the exchange of the four regions separated by dashes.}
\label{Fig3}
\end{figure}

Exact stationary solutions with a non-zero flux can be found from the previous kinetic equations. We usually called these solutions the Kolmogorov-Zakharov spectra. To expose the technique to get such a solution, which implies the Zakharov transformation, I will introduce a simple kinetic equation that must be seen as a toy model because it does not correspond to any physical system. This kinetic equation is the following
\be \label{KEtoy}
\frac{\partial E_k}{\partial t} = \sum_{s_k s_{p} s_{q}} \int_{\Delta} s_{k} \omega_k 
\left[ s_k E_p E_q + s_p E_k E_q + s_q E_k E_p \right] \delta(s_k \omega_k + s_p \omega_p + s_q \omega_q) dp dq \, ,  
\ee
with $\omega_k = k^2$. The kinetic equation is isotropic and $E_k$ is the one dimensional energy spectrum. Also, the integration is limited to a two-dimensional region ($\Delta$) where the triadic relation is satisfied (see Figure \ref{Fig3}). Note that equation (\ref{KEtoy}) verifies the energy conservation. 

%%%%%%%%%%%%%%%%%%%%%%%%%%%%%%%%%%%%%%%%%%%%%%%%%%%%%%
%%%%%%%%%%%%%%%%%%%%%%%%%%%%%%%%%%%%%%%%%%%%%%%%%%%%%%

The first obvious solution that appears is the thermodynamic spectrum, $E_k = A k^{-2}$. Indeed, the substitution of this solution into (\ref{KEtoy}) gives trivially
\be \label{KEtoy2}
\frac{\partial E_k}{\partial t} = \sum_{s_k s_{p} s_{q}} \int_{\Delta} s_{k} \omega_k 
\left[ s_k k^2 + s_p p^2 + s_q q^2 \right] \frac{A^2}{(kpq)^{3}} \delta(s_k \omega_k + s_p \omega_p + s_q \omega_q) dp dq = 0 \, . 
\ee
The second solution is less trivial to obtain. We need to apply the Zakharov transformation.  First, we need to write the previous equation in terms of adimensional wavenumbers. Let us introduce $\xi_p \equiv p/k$ and $\xi_q \equiv q/k$; then, we obtain
\be \label{KEtoy3}
\frac{\partial E_k}{\partial t} = A^2 k^{2+2x} \sum_{s_k s_{p} s_{q}} \int_{\Delta} s_{k} 
\left[ s_k \xi_p^x \xi_q^x + s_p \xi_q^x + s_q \xi_p^x \right] \delta(s_k  + s_p \xi^2_p + s_q \xi^2_q) d\xi_p d\xi_q \, ,  
\ee
where we have also introduced $E_k = A k^x$ with $A$ est positive quantity. The Zakharov transformations correspond to the following change of variables
\be
\xi_{p} \to \frac{1}{\xi_{p}} \, , \quad \xi_{q} \to \frac{\xi_{q}}{\xi_{p}} \, , \quad \rm{(TZ1)} \label{TZ1}
\ee
and
\be
\xi_{p} \to \frac{\xi_{p}}{\xi_{q}} \, , \quad \xi_{q} \to \frac{1}{\xi_{q}} \, . \quad \rm{(TZ2)} \label{TZ2}
\ee
The meaning of these transformations is illustrated in Figure \ref{Fig3}. The method consists of splitting the integral (\ref{KEtoy3}) into three equal contributions,  applying the Zakharov transformation TZ1 to the second and TZ2 to the third, leaving the first integral intact. We then combine the different contributions and, after a few manipulations, obtain the following expression
\ba \label{KEtoy4}
\frac{\partial E_k}{\partial t} &=& \frac{A^2}{3} k^{2+2x} \sum_{s_k s_{p} s_{q}} \int_{\Delta} \xi_q^x \xi_q^x
\left( s_k + s_p \xi_p^{-x} + s_q \xi_q^{-x} \right) \left( s_k + s_p \xi^{-2x-1}_p + s_q \xi^{-2x-1}_q \right) \nonumber \\
&&\delta(s_k + s_p \xi^2_p + s_q \xi^2_q) d\xi_p d\xi_q \, .
\ea
Two solutions appear, $x=-2$ and $x=-3/2$. The first corresponds to the thermodynamic spectrum (as mentioned above) for which the energy flux is zero and the second to the Kolmogorov-Zakharov spectrum with a non-zero energy flux. These properties can be obtained by introducing the isotropic energy flux $\Pi_k$ as follows
\be
\frac{\partial E_k}{\partial t} = - \frac{\partial \Pi_k}{\partial k} = \frac{A^2}{3} k^{2+2x} I(x) \, .
\ee
This leads to the expression
\be
\Pi_k (x) = - \frac{A^2}{3} \frac{k^{3+2x}}{3+2x} I(x) \, .
\ee
We can immediately see that $x=-2$ corresponds to a zero flux since $I(-2)=0$. On the other hand, the value $x=-3/2$ requires a little calculation since $I$ and the denominator are both zero. Using L'Hospital's rule, we find
\be
\varepsilon \equiv \Pi_k (-3/2) = - \frac{A^2}{3} \lim_{x \to -3/2} \frac{I(x)}{3+2x} = \frac{A^2}{3} J \, ,
\ee
with
\ba \label{fluxE}
J \equiv - \lim_{x \to -3/2} \frac{\partial I(x)}{\partial x} &=& \sum_{s_k s_{p} s_{q}} \int_{\Delta} \xi_q^{-3/2} \xi_q^{-3/2}
\left( s_k + s_p \xi_p^{3/2} + s_q \xi_q^{3/2} \right) \left( s_p \xi^{2}_p \ln (\xi_p) + s_q \xi^{2}_q \ln (\xi_q) \right) \nonumber \\
&&\delta(s_k + s_p \xi^2_p + s_q \xi^2_q) d\xi_p d\xi_q \, .
\ea
The direction of the cascade can be obtained numerically from this analytical expression. A positive value for $J$ means a direct energy cascade. The value of $J$ can finally lead to an estimate of the Kolmogorov constant $C_K$ with
\be
E_k = C_K \sqrt{\varepsilon} k^{-3/2} \, \quad \text{and} \quad C_K = \sqrt{\frac{3}{J}} \, .
\ee
The last task is to check the validity of the Kolmogorov-Zakharov spectrum. We require this solution (the inertial range) to be independent of the largest and smallest scales. This is a question of convergence of the collisional integral (\ref{KEtoy3}). Solving this exercise gives a condition $x_1 < x < x_2$ which can prove that our solution $x=-3/2$ is physical if it is inside the domain of convergence. 

In conclusion, we can say that wave turbulence is a solvable problem. Indeed, we have seen from this toy model that in principle we can determine all the properties of wave turbulence by giving the exact solutions, the direction of the cascade, evaluating the Kolmogorov constant and showing the existence of an inertial range. All these properties cannot be found in eddy turbulence, even using the exact Kolmogorov laws.

%%%%%%%%%%%%%%%%%%%%%%%%%%%%%%
\subsection{Applications to Navier-Stokes}

For capillary wave turbulence the isotropic Kolmogorov-Zakharov spectrum is \cite{Zakharov1967}
\be
E_k = A k^{-7/4} \, .
\ee
It is an isotropic spectrum well observed experimentally (see the review \cite{Falcon2022}). 

For inertial wave turbulence the problem is more complex because turbulence is anisotropic with an energy cascade mainly in the direction transverse to the rotation axis. In this case, we have to use a generalization of the Zakharov transformation. This leads to the axisymmetric Kolmogorov-Zakharov spectrum \cite{Galtier2003}
\be
E(k_\perp,k_\parallel) = A k_\perp^{-5/2} k_\parallel^{-1/2} \, .
\ee
Note that the convergence properties and the Kolmogorov constant have been studied only recently \cite{David2023}. This spectrum is observed experimentally and numerically \cite{Monsalve2020,LeReun2020,Yokoyama2021}.

%%%%%%%%%%%%%%%%%%%%%%%%%%%%%%
\begin{appendix}
\section{Appendix: sequential time closures for four-wave interactions}
%%%%%%%%%%%%%%%%%%%%%%%%%%%%%%
\subsection{Fundamental equation}

In this Appendix, we will present a detailed derivation of the kinetic equation for four-wave interactions when we have only cubic nonlinearities. This situation is interesting for example for gravitational waves \cite{Galtier2017}. Consider the following model without dissipation
\be
{\partial u({\bf x},t) \over \partial t} = {\cal L}(u) + \epsilon^2 \, {\cal N}(u,u,u) \, ,
\label{eq1b}
\ee
where $u$ is a random variable such that $\langle u \rangle = 0$, ${\cal L}$ is a linear operator which guarantees that the waves are linear solutions to the problem, and ${\cal N}$ is a non-linear cubic operator (thus involving quartic interactions). The coefficient $\epsilon$ is a small parameter ($0 < \epsilon \ll 1$) which measures the amplitude of the non-linearities (it is originally a measure of the wave amplitude). The Fourier transform of equation (\ref{eq1b}) takes the following schematic form when the canonical variables are introduced
\be \label{66e}
\left( {\partial \over \partial t} + is_k \omega_k \right) A^{s_k} (\kk,t) = 
\epsilon^2 \sum_{s_{p} s_{q} s_r}  \int_{\mathbb{R}^{3d}} \HH_{-\kk \pp \qq \rr} A^{s_p}(\pp,t) A^{s_q}(\qq,t) A^{s_r} (\rr,t) \delta(\kk-\pp-\qq-\rr) d\pp d\qq d\rr \, .
\ee
Here, $\HH_{\kk \pp \qq \rr}$ is an operator symmetric in $\pp$, $\qq$ and $\rr$. 
As for three-wave interactions, we consider a continuous medium which can lead to mathematical difficulties connected with infinite dimensional phase spaces. For this reason, it is preferable to assume a variable spatially periodic over a box of finite size $L$. However, in the derivation of the kinetic equation, the limit $L \to +\infty$ is finally taken (before the long time limit, or equivalently the limit $\epsilon \to 0$). As both approaches lead to the same kinetic equation, for simplicity, we anticipate this result and follow the original approach of \cite{BenneyNewell1967}. 

Equation (\ref{66e}) is simplified by making the following change of variables
\be
A^{s_k}(\kk,t) = a^{s_k}(\kk,t) e^{-i s_k\omega_k t} = a^{s_k}_{k} e^{-i s_k\omega_k t} \, .
\ee
In the interaction representation, we obtain the fundamental equation
\be
{\partial a^{s_k}_{k} \over \partial t} = \epsilon^2 \sum_{s_{p} s_{q} s_r} \int_{\mathbb{R}^{3d}} \HH_{-\kk \pp \qq \rr} a^{s_{p}}_{p} a^{s_{q}}_{q} a^{s_{r}}_{r} 
e^{i \Omega_{k,pqr}t} \delta_{k,pqr} d\pp d\qq d\rr \, ,
\label{eq67}
\ee
with by definition $\delta_{k,pqr} \equiv \delta(\kk-\pp-\qq-\rr)$ and $\Omega_{k,pqr} \equiv s_k\omega_k - s_{p}\omega_p -s_{q}\omega_q-s_{r}\omega_r$. An additional property can be established if we use the relation $A^{s_k}(\kk)=(A^{-s_k}(-\kk))^*$, with $*$ the complex conjugate: then, we have necessarily $\HH_{\kk \pp \qq \rr} = \HH^*_{-\kk -\pp -\qq -\rr}$. Finally, $\HH_{\kk \pp \qq \rr}$ being purely imaginary, we have $\HH^*_{\kk \pp \qq \rr}=-\HH_{\kk \pp \qq \rr}$. This property may be used to simplify the kinetic equations. 

Equation (\ref{eq67}) highlights the temporal evolution of the amplitude of the wave: this evolution is a priori relatively slow since it induces a non-linear term proportional to $\epsilon^2$, which is weaker than for three-wave interactions. The presence of the complex exponential is fundamental for the asymptotic closure: since we are interested in the dynamics over a long time with respect to the period of the waves, the contribution of this exponential is essentially zero. Only some terms will survive: those for which $\Omega_{k,pqr}=0$.  With the condition imposed by the Dirac, we obtain the resonance condition (whose solutions are discussed eg. in \cite{GaltierCUP2023})
\be \label{cr2425}
\left\{
    \begin{array}{ll}
       \kk =  \pp+\qq+\rr \, , \\
	s_k \omega_k =  s_{p}\omega_p +s_{q}\omega_q +s_{r}\omega_r \, .
    \end{array}
\right.
\ee

%%%%%%%%%%%%%%%%%%%%%%%%%%
\subsection{Multiple time scale method}
%%%%%%%%%%%%%%%%%%%%%%%%%%
Following our analysis for three-wave interactions, we use the method of multiple time scales and introduce a sequence of time scales, $T_{0}$, $T_{1}$, $T_{2}$, ..., which will be treated as independent variables, with
\be
T_{0} \equiv t, \quad T_{1}\equiv \epsilon t, \quad T_{2}\equiv \epsilon^{2} t, ... \, .
\ee
The variation of the wave amplitude with $T_1$, $T_2$, $T_3$ and $T_4$ represents the slow variation that we wish to extract. We obtain ($T_0$ being replaced by $t$)
\ba
\left({\partial \over \partial t} + \epsilon {\partial \over \partial T_{1}} + \epsilon^{2} {\partial \over \partial T_{2}} + ...\right) a^{s_k}_{k} =  \epsilon^2 \sum_{s_{p} s_{q} s_r} \int_{\mathbb{R}^{3d}} \HH_{-\kk \pp \qq \rr} a^{s_{p}}_{p} a^{s_{q}}_{q} a^{s_{r}}_{r} e^{i \Omega_{k,pqr}t} \delta_{k,pqr} d\pp d\qq d\rr \, . && 
\label{eq66bis}
\ea
The variable $a^{s_k}_{k}$ must also be expanded to the power of $\epsilon$ and make the various scales appear in time
\be
a^{s_k}_{k} = \sum_{n=0}^{+\infty} \epsilon^n a^{s_k}_{k,n}(t,T_{1},T_{2},...) = a^{s_k}_{k,0} + \epsilon a^{s_k}_{k,1} + \epsilon^{2} a^{s_k}_{k,2} + ... \, .
\ee
%The initial condition will be given only by the leading order amplitude $a^{s}_{k,0}$, while the other lower order contributions will be excited non-linearly. 
%The Gaussian assumption will be made for the initial condition ($t=0$). Unlike for three-wave interaction, this assumption will be used explicitly in the derivation of the kinetic equation. 
The previous expression is then introduced into the fundamental equation (\ref{eq66bis}). We obtain for the first five terms
\begin{subequations}
\begin{align}
{\partial a^{s_k}_{k,0} \over \partial t} &= 0 \, , \\
{\partial a^{s_k}_{k,1} \over \partial t} &= - {\partial a^{s_k}_{k,0} \over \partial T_{1}} \, , \\
{\partial a^{s_k}_{k,2} \over \partial t} &= - {\partial a^{s_k}_{k,1} \over \partial T_{1}} - {\partial a^{s_k}_{k,0} \over \partial T_{2}} + b^{s_k}_{k,2} \, , \\
{\partial a^{s_k}_{k,3} \over \partial t} &= - {\partial a^{s_k}_{k,2} \over \partial T_{1}} - {\partial a^{s_k}_{k,1} \over \partial T_{2}} 
- {\partial a^{s_k}_{k,0} \over \partial T_{3}} + b^{s_k}_{k,3} \\
{\partial a^{s_k}_{k,4} \over \partial t} &= - {\partial a^{s_k}_{k,3} \over \partial T_{1}} - {\partial a^{s_k}_{k,2} \over \partial T_{2}} 
- {\partial a^{s_k}_{k,1} \over \partial T_{3}} - {\partial a^{s_k}_{k,0} \over \partial T_{4}} + b^{s_k}_{k,4} . 
\end{align}
\end{subequations}
with
\be
b^{s_k}_{k,2} = \sum_{s_{p} s_{q} s_r} \int_{\mathbb{R}^{3d}} \HH_{-\kk \pp \qq \rr} a_{p,0}^{s_{p}} a_{q,0}^{s_{q}} a_{r,0}^{s_{r}} e^{i \Omega_{k,pqr}t} \delta_{k,pqr} 
d\pp d\qq d\rr \, , 
\ee
\be
b^{s_k}_{k,3} = \sum_{s_{p} s_{q} s_r} \int_{\mathbb{R}^{3d}} \HH_{-\kk \pp \qq \rr} 
\left[ a_{p,1}^{s_{p}} a_{q,0}^{s_{q}} a_{r,0}^{s_{r}} + a_{p,0}^{s_{p}} a_{q,1}^{s_{q}} a_{r,0}^{s_{r}} + a_{p,0}^{s_{p}} a_{q,0}^{s_{q}} a_{r,1}^{s_{r}} \right]
e^{i \Omega_{k,pqr}t} \delta_{k,pqr} d\pp d\qq d\rr 
\ee
\be
b^{s_k}_{k,4} = \sum_{s_{p} s_{q} s_r} \int_{\mathbb{R}^{3d}} \HH_{-\kk \pp \qq \rr} 
\left[ a_{p,2}^{s_{p}} a_{q,0}^{s_{q}} a_{r,0}^{s_{r}} + a_{p,0}^{s_{p}} a_{q,2}^{s_{q}} a_{r,0}^{s_{r}} + a_{p,0}^{s_{p}} a_{q,0}^{s_{q}} a_{r,2}^{s_{r}} 
+ a_{p,1}^{s_{p}} a_{q,1}^{s_{q}} a_{r,0}^{s_{r}} \right. 
\ee
$$
\left. + a_{p,1}^{s_{p}} a_{q,0}^{s_{q}} a_{r,1}^{s_{r}} + a_{p,0}^{s_{p}} a_{q,1}^{s_{q}} a_{r,1}^{s_{r}} \right] e^{i \Omega_{k,pqr}t} \delta_{k,pqr} d\pp d\qq d\rr \, . 
$$
After integration over $t$, one finds with some simplifications
\begin{subequations}
\begin{align}
a^{s_k}_{k,0} &= a^{s_k}_{k,0}(T_1,T_2,...) \, , \label{137b} \\
a^{s_k}_{k,1} &= -t {\partial a^{s_k}_{k,0} \over \partial T_{1}} \, , \label{138b} \\
a^{s_k}_{k,2} &= \frac{t^2}{2} {\partial^2 a^{s_k}_{k,0} \over \partial T_{1}^2} -t {\partial a^{s_k}_{k,0} \over \partial T_{2}} + \int_0^t b^{s_k}_{k,2} dt \, , \label{139b} \\
a^{s_k}_{k,3} &= -\frac{t^3}{6} {\partial^3 a^{s_k}_{k,0} \over \partial T_{1}^3} + t^2 {\partial^2 a^{s_k}_{k,0} \over \partial T_{1} \partial T_2} 
- t {\partial a^{s_k}_{k,0} \over \partial T_{3}} + \int_0^t b^{s_k}_{k,3} dt - \int_0^t \int_0^t {\partial b^{s_k}_{k,2} \over \partial T_{1}} dt dt \, , \\
a^{s_k}_{k,4} &= \frac{t^4}{24} {\partial^4 a^{s_k}_{k,0} \over \partial T_{1}^4} - \frac{t^3}{2} {\partial^3 a^{s_k}_{k,0} \over \partial T_{1}^2 \partial T_2}
+ t^2 {\partial^2 a^{s_k}_{k,0} \over \partial T_{1} \partial T_3} + \frac{t^2}{2} {\partial^2 a^{s_k}_{k,0} \over \partial T_{2}^2} 
- t {\partial a^{s_k}_{k,0} \over \partial T_{4}} + \int_0^t b^{s_k}_{k,4} dt \\
&- \int_0^t {\partial b^{s_k}_{k,2} \over \partial T_{1}} dt - \int_0^t \int_0^t {\partial b^{s_k}_{k,2} \over \partial T_{2}} dt dt
+ \int_0^t \int_0^t \int_0^t {\partial^2 b^{s_k}_{k,2} \over \partial T_{1}^2} dt dt dt \, . \nonumber
\end{align}
\end{subequations}

%%%%%%%%%%%%%%%%%%%%%%%%%%
\subsection{First asymptotic closure at time $T_1$}
%%%%%%%%%%%%%%%%%%%%%%%%%%

With the previous definitions, the perturbative expansion of the second-order moment writes
\ba \label{37e}
\langle a^{s_k}_{k} a^{s_{k'}}_{k'} \rangle &=& 
\langle (a^{s_k}_{k,0} + \epsilon a^{s_k}_{k,1} + \epsilon^2 a^{s_k}_{k,2} + ...)(a^{s_{k'}}_{k',0} + \epsilon a^{s_{k'}}_{k',1} + \epsilon^2 a^{s_{k'}}_{k',2} + ...) \rangle \\
&=& \langle a^{s_k}_{k,0} a^{s_{k'}}_{k',0} \rangle + \epsilon \langle a^{s_k}_{k,0} a^{s_{k'}}_{k',1} + a^{s_k}_{k,1} a^{s_{k'}}_{k',0} \rangle 
+ \epsilon^2 \langle a^{s_k}_{k,0} a^{s_{k'}}_{k',2} + a^{s_k}_{k,1} a^{s_{k'}}_{k',1} + a^{s_k}_{k,2} a^{s_{k'}}_{k',0}  \rangle + ... \nonumber 
\ea
We assume a turbulence statistically homogeneous, then we have $q^{s_ks_{k'}}(\kk,\kk') \equiv q_{k}^{s_ks_{k'}}$ such that 
\be
\langle a^{s_k}_{k} a^{s_{k'}}_{k'} \rangle = q_{k}^{s_ks_{k'}} \delta (\kk+\kk') \, .
\ee
As with three-wave interactions, we assume that the second-order moment on the left hand side (in fact, the coefficient $q_{k}^{s_ks_{k'}}$ in front of the Dirac function) remains bounded at all time, and so the contributions from the right hand side must also be bounded. At order $\mathcal{O} (\epsilon^0)$, we have the contribution of $\langle a^{s_k}_{k,0} a^{s_{k'}}_{k',0} \rangle$ which will be assumed to be bounded at all times. 

At order $\mathcal{O} (\epsilon^1)$, we find
\be \label{38e4}
\langle a^{s_k}_{k,0} a^{s_{k'}}_{k',1} + a^{s_k}_{k,1} a^{s_{k'}}_{k',0} \rangle = -t \frac{\partial}{\partial T_{1}} \langle a^{s_k}_{k,0} a^{s_{k'}}_{k',0} \rangle \, .
\ee
This unique term is secular which means that over the long time limit we have to impose
\be 
\frac{\partial}{\partial T_{1}} \langle a^{s_k}_{k,0} a^{s_{k'}}_{k',0} \rangle = 0 \, .
\ee
This result can be easily generalized to any nth-order moments. Therefore, the probability density function does not depend on $T_1$ and we can assume that the variable itself does not depend on $T_1$; we have at most
\be
a^{s_k}_{k,0} = a^{s_k}_{k,0}(T_2,T_3,...) \, .
\ee
This result drastically simplifies the previous expressions, which read
\begin{subequations}
\begin{align}
a^{s_k}_{k,1} &= 0 \, , \\
a^{s_k}_{k,2} &= -t {\partial a^{s_k}_{k,0} \over \partial T_{2}} + \int_0^t b^{s_k}_{k,2} dt \, , \\
a^{s_k}_{k,3} &= - t {\partial a^{s_k}_{k,0} \over \partial T_{3}} + \int_0^t b^{s_k}_{k,3} dt \, , \\
a^{s_k}_{k,4} &= \frac{t^2}{2} {\partial^2 a^{s_k}_{k,0} \over \partial T_{2}^2} 
- t {\partial a^{s_k}_{k,0} \over \partial T_{4}} + \int_0^t b^{s_k}_{k,4} dt - \int_0^t \int_0^t {\partial b^{s_k}_{k,2} \over \partial T_{2}} dt dt \, . 
\end{align}
\end{subequations}

%%%%%%%%%%%%%%%%%%%%%%%%%%
\subsection{Second asymptotic closure at time $T_2$}
%%%%%%%%%%%%%%%%%%%%%%%%%%

At order $\mathcal{O} (\epsilon^2)$, we find
\ba
\left\langle a^{s_k}_{k,0} a^{s_{k'}}_{k',2} + a^{s_k}_{k,2} a^{s_{k'}}_{k',0} \right\rangle &=& 
\left\langle a^{s_k}_{k,0} \left(-t {\partial a^{s_{k'}}_{k',0} \over \partial T_{2}} + \int_0^t b^{s_{k'}}_{k',2} dt \right) + 
\left(-t {\partial a^{s_k}_{k,0} \over \partial T_{2}} + \int_0^t b^{s_k}_{k,2} dt \right) a^{s_{k'}}_{k',0} \right\rangle \nonumber \\
&=& -t {\partial \over \partial T_{2}} \langle a^{s_k}_{k,0} a^{s_{k'}}_{k',0} \rangle + 
\left\langle a^{s_k}_{k,0} \int_0^t b^{s_{k'}}_{k',2} dt \right\rangle + \left\langle a^{s_{k'}}_{k',0} \int_0^t b^{s_{k}}_{k,2} dt \right\rangle \, ,
\ea
with for the second term on the right hand side
\be
\left\langle a^{s_k}_{k,0}  \int_0^tb^{s_{k'}}_{k',2} dt \right\rangle = 
\sum_{s_{p} s_{q} s_r} \int_{\mathbb{R}^{3d}} \HH_{-\kk' \pp \qq \rr} \langle a_{p,0}^{s_{p}} a_{q,0}^{s_{q}} a_{r,0}^{s_{r}} a^{s_k}_{k,0} \rangle 
\Delta(\Omega_{k',pqr}) \delta_{k',pqr} d\pp d\qq d\rr \, .
\ee
This term may produce a secular contribution in the long time (recall that $\Delta(0)=t$ and that $\HH_{\kk \pp \qq \rr}$ is purely imaginary) but it is exactly compensated by the third term on the right hand side. Therefore, to keep the moment bounded over the long time, we impose 
\be
{\partial \over \partial T_{2}} \langle a^{s_k}_{k,0} a^{s_{k'}}_{k',0} \rangle = 0 \, .
\ee

%%%%%%%%%%%%%%%%%%%%%%%%%%
\subsection{Third asymptotic closure at time $T_3$}
%%%%%%%%%%%%%%%%%%%%%%%%%%

At next order $\mathcal{O} (\epsilon^3)$, we find
\ba
\left\langle a^{s_k}_{k,0} a^{s_{k'}}_{k',3} + a^{s_k}_{k,3} a^{s_{k'}}_{k',0} \right\rangle &=& 
\left\langle a^{s_k}_{k,0} \left(-t {\partial a^{s_{k'}}_{k',0} \over \partial T_{3}} + \int_0^t b^{s_{k'}}_{k',3} dt \right) + 
\left(-t {\partial a^{s_k}_{k,0} \over \partial T_{3}} + \int_0^t b^{s_k}_{k,3} dt \right) a^{s_{k'}}_{k',0} \right\rangle \nonumber \\
&=& -t {\partial \over \partial T_{3}} \langle a^{s_k}_{k,0} a^{s_{k'}}_{k',0} \rangle + 
\left\langle a^{s_k}_{k,0} \int_0^t b^{s_{k'}}_{k',3} dt + a^{s_{k'}}_{k',0} \int_0^t b^{s_{k}}_{k,3} dt \right\rangle \, ,
\ea
but according to the first closure, $b^{s_{k}}_{k,3}=0$. Therefore, we find the condition over the long time
\be
{\partial \over \partial T_{3}} \langle a^{s_k}_{k,0} a^{s_{k'}}_{k',0} \rangle = 0 \, .
\ee
All these results mean that the dynamics is slower than $T_3$. As we will see later, the kinetic equation for four-wave interactions can be derived to the next order. 

%%%%%%%%%%%%%%%%%%%%%%%%%%
\subsection{Fourth asymptotic closure at time $T_4$}
%%%%%%%%%%%%%%%%%%%%%%%%%%

At order $\mathcal{O} (\epsilon^4)$, we have
\ba 
&&\left\langle a^{s_k}_{k,0} a^{s_{k'}}_{k',4} + a^{s_k}_{k,2} a^{s_{k'}}_{k',2} + a^{s_k}_{k,4} a^{s_{k'}}_{k',0} \right\rangle = \\
&&\left\langle a^{s_k}_{k,0} \left( \frac{t^2}{2} {\partial^2 a^{s_{k'}}_{k',0} \over \partial T_{2}^2} - t {\partial a^{s_{k'}}_{k',0} \over \partial T_{4}} 
+ \int_0^t b^{s_{k'}}_{k',4} dt - \int_0^t \int_0^t {\partial b^{s_{k'}}_{k',2} \over \partial T_{2}} dt dt \right) \right\rangle \nonumber \\
&&+\left\langle a^{s_{k'}}_{k',0} \left( \frac{t^2}{2} {\partial^2 a^{s_{k}}_{k,0} \over \partial T_{2}^2} - t {\partial a^{s_{k}}_{k,0} \over \partial T_{4}} 
+ \int_0^t b^{s_{k}}_{k,4} dt - \int_0^t \int_0^t {\partial b^{s_{k}}_{k,2} \over \partial T_{2}} dt dt \right) \right\rangle \nonumber \\
&&+ \left\langle \left(-t {\partial a^{s_k}_{k,0} \over \partial T_{2}} + \int_0^t b^{s_k}_{k,2} dt\right) 
\left(-t {\partial a^{s_{k'}}_{k',0} \over \partial T_{2}} + \int_0^t b^{s_{k'}}_{k',2} dt\right)\right\rangle \nonumber , 
\ea
which gives after development and simplifications
\be \label{85er}
\left\langle a^{s_k}_{k,0} a^{s_{k'}}_{k',4} + a^{s_k}_{k,2} a^{s_{k'}}_{k',2} + a^{s_k}_{k,4} a^{s_{k'}}_{k',0} \right\rangle = 
\frac{t^2}{2} {\partial^2 \over \partial T_{2}^2} \left\langle a^{s_k}_{k,0} a^{s_{k'}}_{k',0} \right\rangle 
-t {\partial \over \partial T_{4}} \left\langle a^{s_k}_{k,0} a^{s_{k'}}_{k',0} \right\rangle
\ee
$$
+ \left\langle a^{s_{k}}_{k,0} \int_0^t b^{s_{k'}}_{k',4} dt \right\rangle 
+\left\langle a^{s_{k'}}_{k',0} \int_0^t b^{s_{k}}_{k,4} dt \right\rangle 
- \left\langle a^{s_{k}}_{k,0} \int_0^t \int_0^t {\partial b^{s_{k'}}_{k',2} \over \partial T_{2}} dt dt \right\rangle 
- \left\langle a^{s_{k'}}_{k',0}  \int_0^t \int_0^t {\partial b^{s_{k}}_{k,2} \over \partial T_{2}} dt dt \right\rangle
$$
$$
-t \left\langle {\partial a^{s_k}_{k,0} \over \partial T_{2}} \int_0^t b^{s_{k'}}_{k',2} dt \right\rangle 
-t \left\langle {\partial a^{s_{k'}}_{k',0} \over \partial T_{2}} \int_0^t b^{s_k}_{k,2} dt \right\rangle 
+ \left\langle \int_0^t b^{s_k}_{k,2} dt \int_0^t b^{s_{k'}}_{k',2} dt \right\rangle  \, .
$$
According to the second closure, the first term on the right hand side is null. The other terms may give secular contributions. The second term on the right hand side is trivially secular. The third term can be written
\be
\left\langle a^{s_{k}}_{k,0} \int_0^t b^{s_{k'}}_{k',4} dt \right\rangle = 
\ee
$$
\sum_{s_{p} s_{q} s_r} \int_{\mathbb{R}^{3d}} \HH_{-\kk' \pp \qq \rr}  \int_0^t
\left\langle 
a^{s_{k}}_{k,0} a_{p,2}^{s_{p}} a_{q,0}^{s_{q}} a_{r,0}^{s_{r}} 
+ a^{s_{k}}_{k,0} a_{p,0}^{s_{p}} a_{q,2}^{s_{q}} a_{r,0}^{s_{r}} 
+ a^{s_{k}}_{k,0} a_{p,0}^{s_{p}} a_{q,0}^{s_{q}} a_{r,2}^{s_{r}}
\right\rangle 
e^{i \Omega_{k',pqr}t} dt \delta_{k',pqr} d\pp d\qq d\rr \, , 
$$
with
\be
a^{s_k}_{k,2} = -t {\partial a^{s_k}_{k,0} \over \partial T_{2}} + \sum_{s_{p} s_{q} s_r} \int_{\mathbb{R}^{3d}} 
\HH_{-\kk \pp \qq \rr} a_{p,0}^{s_{p}} a_{q,0}^{s_{q}} a_{r,0}^{s_{r}} \Delta(\Omega_{k,pqr}) \delta_{k,pqr} d\pp d\qq d\rr \, .
\ee
We find
\ba
&&\left\langle a^{s_{k}}_{k,0} \int_0^t b^{s_{k'}}_{k',4} dt \right\rangle = 
-\sum_{s_{p} s_{q} s_r} \int_{\mathbb{R}^{3d}} \HH_{-\kk' \pp \qq \rr}  
\left\langle a^{s_{k}}_{k,0} {\partial (a_{p,0}^{s_{p}} a_{q,0}^{s_{q}} a_{r,0}^{s_{r}}) \over \partial T_{2}} \right\rangle 
{\cal W}(\Omega_{k',pqr}) \delta_{k',pqr} d\pp d\qq d\rr \nonumber \\
&&+\sum_{s_{p} s_{q} s_r s_{p'} s_{q'} s_{r'}} \int_{\mathbb{R}^{6d}} 3 \HH_{-\kk' \pp \qq \rr} \HH_{-\pp \pp' \qq' \rr'}
\left\langle a^{s_{k}}_{k,0} a_{q,0}^{s_{q}} a_{r,0}^{s_{r}} a_{p',0}^{s_{p'}} a_{q',0}^{s_{q'}} a_{r',0}^{s_{r'}} \right\rangle 
{\cal X} (\Omega_{k',p'q'r'qr}, \Omega_{k',pqr}) \nonumber \\
&&
\delta_{p,p'q'r'} \delta_{k',pqr}  d\pp' d\qq' d\rr' d\pp d\qq d\rr
\ea
where 
\be
{\cal W}(X) \equiv \int_0^t t e^{i X t} dt = \frac{te^{iXt}}{iX} - \frac{\Delta(X)}{ix} \, , 
\ee
and by definition
\be
{\cal X} (X,Y) \equiv \int_0^t \Delta(X-Y) e^{i Yt} dt = \frac{\Delta(X) - \Delta(Y)}{i(X-Y)} \, ,
\ee
whose long time limit is given by the theory of generalized functions (here we consider $X \neq 0$)
\be
{\cal X} (X,0) \xrightarrow{\text{t $\to +\infty$}} \pi t \delta(X) + it {\cal P} \left(\frac{1}{X}\right) \, .
\ee

The fifth term on the right hand side of expression (\ref{85er}) reads
\ba
&&\left\langle a^{s_{k}}_{k,0} \int_0^t \int_0^t {\partial b^{s_{k'}}_{k',2} \over \partial T_{2}} dt dt \right\rangle = \\
&&
\sum_{s_{p} s_{q} s_r} \int_{\mathbb{R}^{3d}} \HH_{-\kk' \pp \qq \rr} \left\langle a^{s_{k}}_{k,0} 
{\partial (a_{p,0}^{s_{p}} a_{q,0}^{s_{q}} a_{r,0}^{s_{r}}) \over \partial T_{2}} \right\rangle {\cal Y}(\Omega_{k',pqr}) \delta_{k',pqr} d\pp d\qq d\rr \, , \nonumber 
\ea
with by definition
\be
{\cal Y}(X) \equiv \int_0^t \Delta(X) dt = \frac{\Delta(X) - t}{iX} \, . 
\ee

The seventh term on the right hand side of expression (\ref{85er}) reads
\be
t \left\langle {\partial a^{s_k}_{k,0} \over \partial T_{2}} \int_0^t b^{s_{k'}}_{k',2} dt \right\rangle =
t \sum_{s_{p} s_{q} s_r} \int_{\mathbb{R}^{3d}} \HH_{-\kk' \pp \qq \rr}
\left\langle {\partial a^{s_k}_{k,0} \over \partial T_{2}} a_{p,0}^{s_{p}} a_{q,0}^{s_{q}} a_{r,0}^{s_{r}} \right\rangle
\Delta(\Omega_{k',pqr}) \delta_{k',pqr} d\pp d\qq d\rr 
\ee

And the last term on the right hand side of expression (\ref{85er}) reads
\ba
\left\langle \int_0^t b^{s_k}_{k,2} dt \int_0^t b^{s_{k'}}_{k',2} dt \right\rangle &=& 
\sum_{s_{p} s_{q} s_r s_{p'} s_{q'} s_{r'}} \int_{\mathbb{R}^{6d}} \HH_{-\kk \pp \qq \rr} \HH_{-\kk' \pp' \qq' \rr'} 
\left\langle a_{p,0}^{s_{p}} a_{q,0}^{s_{q}} a_{r,0}^{s_{r}} a_{p',0}^{s_{p'}} a_{q',0}^{s_{q'}} a_{r',0}^{s_{r'}} \right\rangle \nonumber \\
&&\Delta(\Omega_{k,pqr}) \Delta(\Omega_{k',p'q'r'}) \delta_{k,pqr} \delta_{k',p'q'r'} d\pp d\qq d\rr d\pp' d\qq' d\rr' \, . 
\ea
We recall long time limit ($X \neq 0$)
\be
\Delta(X) \Delta(-X) \xrightarrow{\text{$t \to +\infty$}} 2 \pi t \delta(X) + 2 {\cal P} \left(\frac{1}{X}\right) \frac{\partial}{\partial X} \, .
\ee

The addition of all these terms gives after simplifications
\ba \label{103e}
&&\left\langle a^{s_k}_{k,0} a^{s_{k'}}_{k',4} + a^{s_k}_{k,2} a^{s_{k'}}_{k',2} + a^{s_k}_{k,4} a^{s_{k'}}_{k',0} \right\rangle = 
-t {\partial \over \partial T_{4}} \left\langle a^{s_k}_{k,0} a^{s_{k'}}_{k',0} \right\rangle \\
&& -t \sum_{s_{p} s_{q} s_r} \int_{\mathbb{R}^{3d}} \HH_{-\kk' \pp \qq \rr}  
{\partial \left\langle a^{s_{k}}_{k,0} a_{p,0}^{s_{p}} a_{q,0}^{s_{q}} a_{r,0}^{s_{r}} \right\rangle \over \partial T_{2}}  
\Delta(\Omega_{k',pqr}) \delta_{k',pqr} d\pp d\qq d\rr \nonumber \\
&&-t \sum_{s_{p} s_{q} s_r} \int_{\mathbb{R}^{3d}} \HH_{-\kk \pp \qq \rr}  
{\partial \left\langle a^{s_{k'}}_{k',0} a_{p,0}^{s_{p}} a_{q,0}^{s_{q}} a_{r,0}^{s_{r}} \right\rangle \over \partial T_{2}}
\Delta(\Omega_{k,pqr}) \delta_{k,pqr} d\pp d\qq d\rr \nonumber \\
&&+\sum_{s_{p} s_{q} s_r s_{p'} s_{q'} s_{r'}} \int_{\mathbb{R}^{6d}} 3 \HH_{-\kk' \pp \qq \rr} \HH_{-\pp \pp' \qq' \rr'}
\left\langle a^{s_{k}}_{k,0} a_{q,0}^{s_{q}} a_{r,0}^{s_{r}} a_{p',0}^{s_{p'}} a_{q',0}^{s_{q'}} a_{r',0}^{s_{r'}} \right\rangle 
{\cal X} (\Omega_{k',p'q'r'qr}, \Omega_{k',pqr}) \nonumber \\
&&\delta_{p,p'q'r'} \delta_{k',pqr}  d\pp' d\qq' d\rr' d\pp d\qq d\rr \nonumber \\
&&+\sum_{s_{p} s_{q} s_r s_{p'} s_{q'} s_{r'}} \int_{\mathbb{R}^{6d}} 3 \HH_{-\kk \pp \qq \rr} \HH_{-\pp \pp' \qq' \rr'}
\left\langle a^{s_{k'}}_{k',0} a_{q,0}^{s_{q}} a_{r,0}^{s_{r}} a_{p',0}^{s_{p'}} a_{q',0}^{s_{q'}} a_{r',0}^{s_{r'}} \right\rangle 
{\cal X} (\Omega_{k,p'q'r'qr}, \Omega_{k,pqr}) \nonumber \\
&&\delta_{p,p'q'r'} \delta_{k,pqr}  d\pp' d\qq' d\rr' d\pp d\qq d\rr \nonumber \\
&&+\sum_{s_{p} s_{q} s_r s_{p'} s_{q'} s_{r'}} \int_{\mathbb{R}^{6d}} \HH_{-\kk \pp \qq \rr} \HH_{-\kk' \pp' \qq' \rr'} 
\left\langle a_{p,0}^{s_{p}} a_{q,0}^{s_{q}} a_{r,0}^{s_{r}} a_{p',0}^{s_{p'}} a_{q',0}^{s_{q'}} a_{r',0}^{s_{r'}} \right\rangle 
\Delta(\Omega_{k,pqr}) \Delta(\Omega_{k',p'q'r'}) \nonumber \\
&&\delta_{k,pqr} \delta_{k',p'q'r'} d\pp d\qq d\rr d\pp' d\qq' d\rr' \, . \nonumber
\ea
To further simplify the previous expression one needs to consider the fourth-order moment. With $a^{s_k}_{k,1} = 0$, we find
\ba 
\langle a^{s_k}_{k} a^{s_{k'}}_{k'} a^{s_{k''}}_{k''} a^{s_{k'''}}_{k'''}  \rangle &=& 
\langle (a^{s_k}_{k,0} + \epsilon^2 a^{s_k}_{k,2} + ...)(a^{s_{k'}}_{k',0} + \epsilon^2 a^{s_{k'}}_{k',2} + ...) ... \rangle \\
&=& \langle a^{s_k}_{k,0} a^{s_{k'}}_{k',0} ... \rangle 
+ \epsilon^2 \langle a^{s_k}_{k,2} a^{s_{k'}}_{k',0}... + a^{s_k}_{k,0} a^{s_{k'}}_{k',2}... + ... \rangle + ... \nonumber 
\ea
As before, we assume that the left hand side is bounded, which means that at each order the contribution from the right hand side must also be bounded. At order $\mathcal{O} (\epsilon^2)$ we have
\ba
&&\langle a^{s_k}_{k,2} a^{s_{k'}}_{k',0} a^{s_{k''}}_{k'',0} a^{s_{k'''}}_{k''',0}  + a^{s_k}_{k,0} a^{s_{k'}}_{k',2} a^{s_{k''}}_{k'',0} a^{s_{k'''}}_{k''',0}  + ... \rangle = \\
&&\left\langle \left(-t {\partial a^{s_k}_{k,0} \over \partial T_{2}} + \int_0^t b^{s_k}_{k,2} dt \right) a^{s_{k'}}_{k',0} a^{s_{k''}}_{k'',0} a^{s_{k'''}}_{k''',0} 
+ a^{s_k}_{k,0} \left(-t {\partial a^{s_{k'}}_{k',0} \over \partial T_{2}} + \int_0^t b^{s_{k'}}_{k',2} dt \right)a^{s_{k''}}_{k'',0} a^{s_{k'''}}_{k''',0} 
+ ... \right\rangle \nonumber \\
&&
= -t {\partial \left\langle a^{s_k}_{k,0} a^{s_{k'}}_{k',0} a^{s_{k''}}_{k'',0} a^{s_{k'''}}_{k''',0}  \right\rangle \over \partial T_2} \nonumber \\
&&+ \sum_{s_{p} s_{q} s_r} \int_{\mathbb{R}^{3d}} \HH_{-\kk \pp \qq \rr} 
\left\langle  a_{p,0}^{s_{p}} a_{q,0}^{s_{q}} a_{r,0}^{s_{r}} a^{s_{k'}}_{k',0} a^{s_{k''}}_{k'',0} a^{s_{k'''}}_{k''',0} \right\rangle
\Delta(\Omega_{k,pqr}) \delta_{k,pqr} d\pp d\qq d\rr + ... \, .
\nonumber
\ea
Secular contributions (with $\Delta(0) = t$) can emerge from terms involving sixth-order moments which must be cancelled with the first term in the right hand side. We obtain three types of contribution, namely
\ba
&&{\partial \left\langle a^{s_k}_{k,0} a^{s_{k'}}_{k',0} a^{s_{k''}}_{k'',0} a^{s_{k'''}}_{k''',0}  \right\rangle \over \partial T_2} = 
\sum_{s_p} \int_{\mathbb{R}^{d}} \HH_{-\kk \pp -\pp \kk} q_{p,0}^{s_p -s_p} 
\left\langle a_{k,0}^{s_{k}} a^{s_{k'}}_{k',0} a^{s_{k''}}_{k'',0} a^{s_{k'''}}_{k''',0} \right\rangle d\pp \\
&&
+ \sum_{s_p} \int_{\mathbb{R}^{d}} \HH_{-\kk \pp \kk -\pp} q_{p,0}^{s_p -s_p} 
\left\langle a_{k,0}^{s_{k}} a^{s_{k'}}_{k',0} a^{s_{k''}}_{k'',0} a^{s_{k'''}}_{k''',0} \right\rangle d\pp 
+ \sum_{s_q} \int_{\mathbb{R}^{d}} \HH_{-\kk \kk \qq -\qq} q_{q,0}^{s_q -s_q} 
\left\langle a_{k,0}^{s_{k}} a^{s_{k'}}_{k',0} a^{s_{k''}}_{k'',0} a^{s_{k'''}}_{k''',0} \right\rangle d\qq  + ... .
\nonumber
\ea
Using the symmetries of $\HH_{\kk \pp \qq \rr}$, we eventually find
\be
{\partial \left\langle a^{s_k}_{k,0} a^{s_{k'}}_{k',0} a^{s_{k''}}_{k'',0} a^{s_{k'''}}_{k''',0}  \right\rangle \over \partial T_2} = 
\ee
$$3 \left\langle a_{k,0}^{s_{k}} a^{s_{k'}}_{k',0} a^{s_{k''}}_{k'',0} a^{s_{k'''}}_{k''',0} \right\rangle
\sum_{s_p}  \int_{\mathbb{R}^{d}} \left( \HH_{-\kk \kk -\pp \pp} + \HH_{-\kk' \kk' -\pp \pp} + \HH_{-\kk'' \kk'' -\pp \pp} + \HH_{-\kk''' \kk'''' -\pp \pp} \right)
q_{p,0}^{s_p -s_p} d\pp  .
$$
From the second closure, we know that the second-order moment does not depend on $T_2$, and therefore the left hand side can be reduced to a fourth-order cumulant. On the right hand side, we see that the decomposition in terms of second-order moments does not give any contribution because of the pairwise cancellation of the $\HH$ coefficients. Therefore, this equation can be reduced to (with $\kk+\kk'+\kk''+\kk'''={\bf 0}$)
\be
{\partial q_{kk'k''}^{s_k s_{k'} s_{k''} s_{k'''}} \over \partial T_2}  = 
\ee
$$
3 q_{kk'k''}^{s_k s_{k'} s_{k''} s_{k'''}} 
\sum_{s_p}  \int_{\mathbb{R}^{d}} \left( \HH_{-\kk \kk -\pp \pp} + \HH_{-\kk' \kk' -\pp \pp} + \HH_{-\kk'' \kk'' -\pp \pp} + \HH_{-\kk''' \kk'''' -\pp \pp} \right)
q_{p,0}^{s_p -s_p} d\pp  
$$
whose exact solution is
\be
q_{kk'k''}^{s_k s_{k'} s_{k''} s_{k'''}} = q_0 e^{3 T_2 \sum_{s_p}  \int_{\mathbb{R}^{d}} 
\left( \HH_{-\kk \kk -\pp \pp} + \HH_{-\kk' \kk' -\pp \pp} + \HH_{-\kk'' \kk'' -\pp \pp} + \HH_{-\kk''' \kk'''' -\pp \pp} \right)
q_{p,0}^{s_p -s_p} d\pp} \, , \label{120}
\ee
with $q_0 = q_{kk'k''}^{s_k s_{k'} s_{k''} s_{k'''}}(t=0)$ the value of the fourth-order cumulant at $t=0$. 
Also, if the fourth-order cumulant is zero initially 
(if we assume a Gaussian statistics), 
%(if we assume the random phase approximation), 
it will remain so at time $T_2$, and there is therefore no contribution from the derivatives in $T_2$ in expression (\ref{103e}).  
This property may be linked to the propagation of chaos discussed by mathematicians \cite{Deng2023}. 
Note that the assumption of random phase approximation is often made for the derivation of the kinetic equation \cite{Nazarenko11}. 
We can also note that the solution (\ref{120}) corresponds to an oscillating term (we recall that $\HH_{\kk \pp \qq \rr}$ is purely imaginary and $q_{p,0}^{s_p -s_p}$ is real) and keeping this term in the kinetic equation will not lead to a contribution at time $T_4$.
Therefore, we obtain
\ba \label{110e}
&&\left\langle a^{s_k}_{k,0} a^{s_{k'}}_{k',4} + a^{s_k}_{k,2} a^{s_{k'}}_{k',2} + a^{s_k}_{k,4} a^{s_{k'}}_{k',0} \right\rangle = 
-t {\partial \over \partial T_{4}} \left\langle a^{s_k}_{k,0} a^{s_{k'}}_{k',0} \right\rangle \\
&&+\sum_{s_{p} s_{q} s_r s_{p'} s_{q'} s_{r'}} \int_{\mathbb{R}^{6d}} \HH_{-\kk \pp \qq \rr} \HH_{-\kk' \pp' \qq' \rr'} 
\left\langle a_{p,0}^{s_{p}} a_{q,0}^{s_{q}} a_{r,0}^{s_{r}} a_{p',0}^{s_{p'}} a_{q',0}^{s_{q'}} a_{r',0}^{s_{r'}} \right\rangle 
\Delta(\Omega_{k,pqr}) \Delta(\Omega_{k',p'q'r'}) \nonumber \\
&&\delta_{k,pqr} \delta_{k',p'q'r'} d\pp d\qq d\rr d\pp' d\qq' d\rr' \nonumber \\
&&+3 \sum_{s_{p} s_{q} s_r s_{p'} s_{q'} s_{r'}} \int_{\mathbb{R}^{6d}} \HH_{-\kk' \pp \qq \rr} \HH_{-\pp \pp' \qq' \rr'}
\left\langle a^{s_{k}}_{k,0} a_{q,0}^{s_{q}} a_{r,0}^{s_{r}} a_{p',0}^{s_{p'}} a_{q',0}^{s_{q'}} a_{r',0}^{s_{r'}} \right\rangle 
{\cal X} (\Omega_{k',p'q'r'qr}, \Omega_{k',pqr}) \nonumber \\
&&\delta_{p,p'q'r'} \delta_{k',pqr}  d\pp' d\qq' d\rr' d\pp d\qq d\rr \nonumber \\
&&+3 \sum_{s_{p} s_{q} s_r s_{p'} s_{q'} s_{r'}} \int_{\mathbb{R}^{6d}} \HH_{-\kk \pp \qq \rr} \HH_{-\pp \pp' \qq' \rr'}
\left\langle a^{s_{k'}}_{k',0} a_{q,0}^{s_{q}} a_{r,0}^{s_{r}} a_{p',0}^{s_{p'}} a_{q',0}^{s_{q'}} a_{r',0}^{s_{r'}} \right\rangle 
{\cal X} (\Omega_{k,p'q'r'qr}, \Omega_{k,pqr}) \nonumber \\
&&\delta_{p,p'q'r'} \delta_{k,pqr}  d\pp' d\qq' d\rr' d\pp d\qq d\rr  \, , \nonumber
\ea
from which it remains to find the secular contributions. To do this, it is necessary to develop the sixth-order moments into the product of the lower order moments plus the cumulant. 

Let us illustrate this with the second term on the right hand side of the previous expression. We have 41 terms which are written (to simplify the writing, we keep the moment notation of the fourth-order cumulant; we only use the cumulant notation for the sixth-order cumulant for which there is an ambiguity)
\be \label{decompc}
\langle a_{p,0}^{s_{p}} a_{q,0}^{s_{q}} a_{r,0}^{s_{r}} a_{p',0}^{s_{p'}} a_{q',0}^{s_{q'}} a_{r',0}^{s_{r'}} \rangle = 
\ee
$$
\langle a_{p,0}^{s_{p}} a_{q,0}^{s_{q}} \rangle \langle a_{r,0}^{s_{r}} a_{p',0}^{s_{p'}} \rangle \langle a_{q',0}^{s_{q'}} a_{r',0}^{s_{r'}} \rangle
+ \langle a_{p,0}^{s_{p}} a_{r,0}^{s_{r}} \rangle \langle a_{q,0}^{s_{q}} a_{p',0}^{s_{p'}} \rangle \langle a_{q',0}^{s_{q'}} a_{r',0}^{s_{r'}} \rangle
+ \langle a_{p,0}^{s_{p}} a_{p',0}^{s_{p'}} \rangle \langle a_{q,0}^{s_{q}} a_{r,0}^{s_{r}} \rangle \langle a_{q',0}^{s_{q'}} a_{r',0}^{s_{r'}} \rangle
$$
$$
+\langle a_{p,0}^{s_{p}} a_{q',0}^{s_{q'}} \rangle \langle a_{q,0}^{s_{q}} a_{r,0}^{s_{r}} \rangle \langle a_{p',0}^{s_{p'}} a_{r',0}^{s_{r'}} \rangle
+ \langle a_{p,0}^{s_{p}} a_{r',0}^{s_{r'}} \rangle \langle a_{q,0}^{s_{q}} a_{r,0}^{s_{r}} \rangle \langle a_{p',0}^{s_{p'}} a_{q',0}^{s_{q'}} \rangle
+ \langle a_{p,0}^{s_{p}} a_{q,0}^{s_{q}} \rangle \langle a_{r,0}^{s_{r}} a_{q',0}^{s_{q'}} \rangle \langle a_{p',0}^{s_{p'}} a_{r',0}^{s_{r'}} \rangle
$$
$$
+\langle a_{p,0}^{s_{p}} a_{q,0}^{s_{q}} \rangle \langle a_{r,0}^{s_{r}} a_{r',0}^{s_{r'}} \rangle \langle a_{p',0}^{s_{p'}} a_{q',0}^{s_{q'}} \rangle
+ \langle a_{p,0}^{s_{p}} a_{r,0}^{s_{r}} \rangle \langle a_{q,0}^{s_{q}} a_{r',0}^{s_{r'}} \rangle \langle a_{p',0}^{s_{p'}} a_{q',0}^{s_{q'}} \rangle
+ \langle a_{p,0}^{s_{p}} a_{r,0}^{s_{r}} \rangle \langle a_{q,0}^{s_{q}} a_{q',0}^{s_{q'}} \rangle \langle a_{p',0}^{s_{p'}} a_{r',0}^{s_{r'}} \rangle
$$
$$
+\langle a_{p,0}^{s_{p}} a_{p',0}^{s_{p'}} \rangle \langle a_{q,0}^{s_{q}} a_{q',0}^{s_{q'}} \rangle \langle a_{r,0}^{s_{r}} a_{r',0}^{s_{r'}} \rangle
+ \langle a_{p,0}^{s_{p}} a_{p',0}^{s_{p'}} \rangle \langle a_{q,0}^{s_{q}} a_{r',0}^{s_{r'}} \rangle \langle a_{r,0}^{s_{r}} a_{q',0}^{s_{q'}} \rangle
+ \langle a_{p,0}^{s_{p}} a_{q',0}^{s_{q'}} \rangle \langle a_{q,0}^{s_{q}} a_{p',0}^{s_{p'}} \rangle \langle a_{r,0}^{s_{r}} a_{r',0}^{s_{r'}} \rangle
$$
$$
+\langle a_{p,0}^{s_{p}} a_{q',0}^{s_{q'}} \rangle \langle a_{q,0}^{s_{q}} a_{p',0}^{s_{p'}} \rangle \langle a_{r,0}^{s_{r}} a_{p',0}^{s_{p'}} \rangle
+ \langle a_{p,0}^{s_{p}} a_{r',0}^{s_{r'}} \rangle \langle a_{q,0}^{s_{q}} a_{p',0}^{s_{p'}} \rangle \langle a_{r,0}^{s_{r}} a_{q',0}^{s_{q'}} \rangle
+ \langle a_{p,0}^{s_{p}} a_{r',0}^{s_{r'}} \rangle \langle a_{q,0}^{s_{q}} a_{q',0}^{s_{q'}} \rangle \langle a_{r,0}^{s_{r}} a_{r',0}^{s_{r'}} \rangle
$$
$$
+\langle a_{p,0}^{s_{p}} a_{q,0}^{s_{q}} a_{r,0}^{s_{r}} \rangle \langle a_{p',0}^{s_{p'}} a_{q',0}^{s_{q'}} a_{r',0}^{s_{r'}} \rangle
+\langle a_{p,0}^{s_{p}} a_{q,0}^{s_{q}} a_{p',0}^{s_{p'}} \rangle \langle a_{r,0}^{s_{r}} a_{q',0}^{s_{q'}} a_{r',0}^{s_{r'}} \rangle
+\langle a_{p,0}^{s_{p}} a_{q,0}^{s_{q}} a_{q',0}^{s_{q'}} \rangle \langle a_{r,0}^{s_{r}} a_{p',0}^{s_{p'}} a_{r',0}^{s_{r'}} \rangle
$$
$$
+\langle a_{p,0}^{s_{p}} a_{q,0}^{s_{q}} a_{r',0}^{s_{r'}} \rangle \langle a_{r,0}^{s_{r}} a_{p',0}^{s_{p'}} a_{q',0}^{s_{q'}} \rangle
+\langle a_{p,0}^{s_{p}} a_{p',0}^{s_{p'}} a_{r,0}^{s_{r}} \rangle \langle a_{q,0}^{s_{q}} a_{q',0}^{s_{q'}} a_{r',0}^{s_{r'}} \rangle
+\langle a_{p,0}^{s_{p}} a_{q',0}^{s_{q'}} a_{r',0}^{s_{r'}} \rangle \langle a_{p',0}^{s_{p'}} a_{q',0}^{s_{q'}} a_{r,0}^{s_{r}} \rangle
$$
$$
+\langle a_{p',0}^{s_{p'}} a_{q,0}^{s_{q}} a_{r,0}^{s_{r}} \rangle \langle a_{p,0}^{s_{p}} a_{q',0}^{s_{q'}} a_{r',0}^{s_{r'}} \rangle
+\langle a_{q',0}^{s_{q'}} a_{q,0}^{s_{q}} a_{r,0}^{s_{r}} \rangle \langle a_{p',0}^{s_{p'}} a_{p,0}^{s_{p}} a_{r',0}^{s_{r'}} \rangle
+\langle a_{r',0}^{s_{r'}} a_{q,0}^{s_{q}} a_{r,0}^{s_{r}} \rangle \langle a_{p',0}^{s_{p'}} a_{q',0}^{s_{q'}} a_{p,0}^{s_{p}} \rangle
$$
$$
+\langle a_{p,0}^{s_{p}} a_{r',0}^{s_{r'}} a_{r,0}^{s_{r}} \rangle \langle a_{q,0}^{s_{q}} a_{q',0}^{s_{q'}} a_{p',0}^{s_{p'}} \rangle
$$
$$
+\langle a_{p,0}^{s_{p}} a_{q,0}^{s_{q}} \rangle \langle a_{r,0}^{s_{r}} a_{p',0}^{s_{p'}} a_{q',0}^{s_{q'}} a_{r',0}^{s_{r'}} \rangle
+\langle a_{p,0}^{s_{p}} a_{r,0}^{s_{r}} \rangle \langle a_{q,0}^{s_{q}} a_{p',0}^{s_{p'}} a_{q',0}^{s_{q'}} a_{r',0}^{s_{r'}} \rangle
+\langle a_{p,0}^{s_{p}} a_{p',0}^{s_{p'}} \rangle \langle a_{r,0}^{s_{r}} a_{q,0}^{s_{q}} a_{q',0}^{s_{q'}} a_{r',0}^{s_{r'}} \rangle
$$
$$
+\langle a_{p,0}^{s_{p}} a_{q',0}^{s_{q'}} \rangle \langle a_{r,0}^{s_{r}} a_{q,0}^{s_{q}}  a_{p',0}^{s_{p'}} a_{r',0}^{s_{r'}} \rangle
+\langle a_{p,0}^{s_{p}} a_{r',0}^{s_{r'}} \rangle \langle a_{q,0}^{s_{q}} a_{r,0}^{s_{r}} a_{p',0}^{s_{p'}} a_{q',0}^{s_{q'}} \rangle
+\langle a_{r,0}^{s_{r}} a_{q,0}^{s_{q}} \rangle \langle a_{p,0}^{s_{p}} a_{p',0}^{s_{p'}} a_{q',0}^{s_{q'}} a_{r',0}^{s_{r'}} \rangle
$$
$$
+\langle a_{p',0}^{s_{p'}} a_{q,0}^{s_{q}} \rangle \langle a_{r,0}^{s_{r}} a_{p,0}^{s_{p}}  a_{q',0}^{s_{q'}} a_{r',0}^{s_{r'}} \rangle
+\langle a_{q,0}^{s_{q}} a_{q',0}^{s_{q'}} \rangle \langle a_{r,0}^{s_{r}} a_{p,0}^{s_{}} a_{p',0}^{s_{p'}} a_{r',0}^{s_{r'}} \rangle
+\langle a_{q,0}^{s_{q}} a_{r',0}^{s_{r'}} \rangle \langle a_{p,0}^{s_{p}} a_{r,0}^{s_{r}} a_{p',0}^{s_{p'}} a_{q',0}^{s_{q'}} \rangle
$$
$$
+\langle a_{r,0}^{s_{r}} a_{p',0}^{s_{p'}} \rangle \langle a_{p,0}^{s_{p}} a_{q,0}^{s_{q}}  a_{q',0}^{s_{q'}} a_{r',0}^{s_{r'}} \rangle
+\langle a_{r,0}^{s_{r}} a_{q',0}^{s_{q'}} \rangle \langle a_{p,0}^{s_{p}} a_{q,0}^{s_{q}} a_{p',0}^{s_{p'}} a_{r',0}^{s_{r'}} \rangle
+\langle a_{r,0}^{s_{r}} a_{r',0}^{s_{r'}} \rangle \langle a_{p,0}^{s_{p}} a_{q,0}^{s_{q}} a_{p',0}^{s_{p'}} a_{q',0}^{s_{q'}} \rangle
$$
$$
+\langle a_{p',0}^{s_{p'}} a_{q',0}^{s_{q'}} \rangle \langle a_{p,0}^{s_{p}} a_{q,0}^{s_{q}}  a_{r,0}^{s_{r}} a_{r',0}^{s_{r'}} \rangle
+\langle a_{p',0}^{s_{p'}} a_{r',0}^{s_{r'}} \rangle \langle a_{p,0}^{s_{p}} a_{q,0}^{s_{q}} a_{p',0}^{s_{p'}} a_{r,0}^{s_{r}} \rangle
+\langle a_{q',0}^{s_{q'}} a_{r',0}^{s_{r'}} \rangle \langle a_{p,0}^{s_{p}} a_{q,0}^{s_{q}} a_{p',0}^{s_{p'}} a_{r,0}^{s_{r}} \rangle
$$
$$
+q_{pqrp'q'}^{s_p s_q s_r s_{p'} s_{q'} s_{r'}} \delta(\pp+\qq+\rr+\pp'+\qq'+\rr') \, .
$$
The same form is obtained for the third and fourth terms of expression (\ref{110e}). 

We will consider the super-secular contributions $\propto t^2$ that can emerge because of the relation $\Delta(0) \Delta(0) = t^2$. 
The first nine terms give the same super-secular contributions; after simplification one finds
\be
9 t^2 \sum_{s_{p} s_q} \int_{\mathbb{R}^{2d}} \HH_{\kk \kk \pp -\pp} \HH_{-\kk -\kk \qq -\qq} q_k^{s_k} q_p^{s_p} q_q^{s_q} \delta_{kk'} d\pp d\qq \, .
\ee
The third term in expression (\ref{110e}) gives the following super-secular contributions (due to ${\cal X} (0,0) = t^2/2$)
\be
\frac{9 t^2}{2} \sum_{s_{p} s_q} \int_{\mathbb{R}^{2d}} \HH_{\kk \kk \pp -\pp} \HH_{\kk \kk \qq -\qq} q_k^{s_k} q_p^{s_p} q_q^{s_q} \delta_{kk'} d\pp d\qq \, .
\ee
Finally, the fourth term in expression (\ref{110e}) gives for the same reason the same super-secular contributions as above. The addition of the three super-secular contributions gives zero because of the relation $\HH_{-\kk -\kk \qq -\qq}= -\HH_{\kk \kk -\qq \qq} = -\HH_{\kk \kk \qq -\qq}$. 

Next, we will consider the (classical) secular terms $\propto t$. The first nine terms of expression (\ref{decompc}) give no contribution. The next six terms give the same contribution, the addition of which gives
\be
12 \pi t \sum_{s_{p} s_{q} s_r} \int_{\mathbb{R}^{3d}} \HH_{\kk \pp \qq \rr} \HH_{-\kk -\pp -\qq -\rr} q_p^{s_p} q_q^{s_q} q_r^{s_r}
\delta_{k,pqr} \delta(\Omega_{k,pqr}) d\pp d\qq d\rr \delta_{kk'} \, .
\ee

The product of the third-order cumulant give no secular contribution, but the six products of the second-order cumulant with fourth-order cumulant give a secular contribution. We find the following expression
\ba
3 t \sum_{s_{p} s_{p'} s_{q'} s_{r'}} \int_{\mathbb{R}^{4d}} \HH_{\kk \kk \pp -\pp} \HH_{-\kk \pp' \qq' \rr'} 
q_p^{s_p} q_{p'q'r'}^{s_{r} s_{p'} s_{q'} s_{r'}} \delta_{-k,p'q'r'} 
\Delta(\Omega_{-k,p'q'r'}) 
%\left( \pi \delta(\Omega_{-k,p'q'r'}) -i {\cal P} \left( \frac{1}{\Omega_{-k,p'q'r'}} \right) \right)
d\pp d\pp' d\qq' d\rr' \delta_{kk'} \\
+3 t \sum_{s_{p} s_{q} s_{r} s_{p'}} \int_{\mathbb{R}^{4d}} \HH_{\kk \pp \qq \rr} \HH_{-\kk -\kk \pp' -\pp'} 
q_{p'}^{s_{p'}} q_{pqr}^{s_{p} s_{q} s_{r} s_{r'}} \delta_{k,pqr} \Delta(\Omega_{k,pqr}) d\pp d\qq d\rr d\pp' \delta_{kk'}  \, , \nonumber
\ea
for which the long time limit is given by the Riemann-Lebesgue lemma. Finally, the sixth-order cumulant gives no secular contribution. 

The third term in expression (\ref{110e}) gives secular contributions. First of all, they come from products of second-order cumulants; we obtain in the long time limit
\be
18t \sum_{s_{p} s_{q} s_{r}} \int_{\mathbb{R}^{3d}} \HH_{\kk \pp \qq \rr} \HH_{\pp \kk -\qq -\rr} q_k^{s_k} q_q^{s_q} q_r^{s_r}  \delta_{k,pqr} 
\left( \pi \delta(\Omega_{k,pqr}) -i {\cal P} \left( \frac{1}{\Omega_{k,pqr}} \right) \right) d\pp d\qq d\rr \delta_{kk'} 
\, . \nonumber
\ee
The other contributions come from products of second-order cumulant with fourth-order cumulant; we have 
\ba
6 \sum_{s_{p} s_{p'} s_{q'} s_{r'}} \int_{\mathbb{R}^{4d}} \HH_{\kk \kk \pp -\pp} \HH_{\pp \pp' \qq' \rr'} 
q_k^{s_k} q_{p'q'r'}^{s_{r} s_{p'} s_{q'} s_{r'}} \delta_{p,p'q'r'} {\cal X} (\Omega_{p,p'q'r'},0) d\pp d\pp' d\qq' d\rr' \delta_{kk'} \\
+3 \sum_{s_{r} s_{p'} s_{q'} s_{r'}} \int_{\mathbb{R}^{4d}} \HH_{\kk \kk \rr -\rr} \HH_{\kk \pp' \qq' \rr'} 
q_r^{s_r} q_{p'q'r'}^{s_{k'} s_{p'} s_{q'} s_{r'}} \delta_{k,p'q'r'} 
{\cal X} (\Omega_{k,p'q'r'},0) d\rr d\pp' d\qq' d\rr' \delta_{kk'} \, , \nonumber
\ea
with the long time limit
\be
{\cal X} (X,0) \xrightarrow{\text{t $\to +\infty$}} \pi t \delta(X) + it {\cal P} \left(\frac{1}{X}\right) \, .
\ee
The fourth term in expression (\ref{110e}) gives secular contributions similar to the third term (we only need to exchange $\kk$ and $\kk'$). 

Then, we obtain the following condition to cancel the contribution of secular terms in the long time limit
\be
{\partial q_k^{s_k} \over \partial T_{4}} = 
12 \pi \sum_{s_{p} s_{q} s_r} \int_{\mathbb{R}^{3d}} \HH_{\kk \pp \qq \rr} \HH_{-\kk -\pp -\qq -\rr} q_p^{s_p} q_q^{s_q} q_r^{s_r}
\delta_{k,pqr} \delta(\Omega_{k,pqr}) d\pp d\qq d\rr
\ee
$$
+3 \sum_{s_{p} s_{p'} s_{q'} s_{r'}} \int_{\mathbb{R}^{4d}} \HH_{\kk \kk \pp -\pp} \HH_{-\kk \pp' \qq' \rr'} 
q_p^{s_p} q_{p'q'r'}^{s_{r} s_{p'} s_{q'} s_{r'}} \delta_{-k,p'q'r'} 
\left( \pi \delta(\Omega_{-k,p'q'r'}) -i {\cal P} \left( \frac{1}{\Omega_{-k,p'q'r'}} \right) \right) d\pp d\pp' d\qq' d\rr'
$$
$$+3 \sum_{s_{p} s_{q} s_{r} s_{p'}} \int_{\mathbb{R}^{4d}} \HH_{\kk \pp \qq \rr} \HH_{-\kk -\kk \pp' -\pp'} 
q_{p'}^{s_{p'}} q_{pqr}^{s_{p} s_{q} s_{r} s_{r'}} \delta_{k,pqr} 
\left( \pi \delta(\Omega_{k,pqr}) -i {\cal P} \left( \frac{1}{\Omega_{k,pqr}} \right) \right) d\pp d\qq d\rr d\pp'
$$
$$
+18 \sum_{s_{p} s_{q} s_{r}} \int_{\mathbb{R}^{3d}} \HH_{\kk \pp \qq \rr} \HH_{\pp \kk -\qq -\rr} q_k^{s_k} q_q^{s_q} q_r^{s_r}  \delta_{k,pqr} 
\left( \pi \delta(\Omega_{k,pqr}) -i {\cal P} \left( \frac{1}{\Omega_{k,pqr}} \right) \right) d\pp d\qq d\rr
$$
$$
+6 \sum_{s_{p} s_{p'} s_{q'} s_{r'}} \int_{\mathbb{R}^{4d}} \HH_{\kk \kk \pp -\pp} \HH_{\pp \pp' \qq' \rr'} 
q_k^{s_k} q_{p'q'r'}^{s_{r} s_{p'} s_{q'} s_{r'}} \delta_{p,p'q'r'} 
\left( \pi \delta(\Omega_{p,p'q'r'}) -i {\cal P} \left( \frac{1}{\Omega_{p,p'q'r'}} \right) \right) 
d\pp d\pp' d\qq' d\rr'
$$
$$
+3 \sum_{s_{r} s_{p'} s_{q'} s_{r'}} \int_{\mathbb{R}^{4d}} \HH_{\kk \kk \rr -\rr} \HH_{\kk \pp' \qq' \rr'} 
q_r^{s_r} q_{p'q'r'}^{s_{k'} s_{p'} s_{q'} s_{r'}} \delta_{k,p'q'r'} 
\left( \pi \delta(\Omega_{k,p'q'r'}) -i {\cal P} \left( \frac{1}{\Omega_{k,p'q'r'}} \right) \right) 
d\rr d\pp' d\qq' d\rr'
$$
$$
+18 \sum_{s_{p} s_{q} s_{r}} \int_{\mathbb{R}^{3d}} \HH_{-\kk \pp \qq \rr} \HH_{\pp -\kk -\qq -\rr} q_k^{s_k} q_q^{s_q} q_r^{s_r}  \delta_{-k,pqr} 
\left( \pi \delta(\Omega_{-k,pqr}) -i {\cal P} \left( \frac{1}{\Omega_{-k,pqr}} \right) \right) d\pp d\qq d\rr
$$
$$
+6 \sum_{s_{p} s_{p'} s_{q'} s_{r'}} \int_{\mathbb{R}^{4d}} \HH_{-\kk -\kk \pp -\pp} \HH_{\pp \pp' \qq' \rr'} 
q_k^{s_k} q_{p'q'r'}^{s_{r} s_{p'} s_{q'} s_{r'}} \delta_{p,p'q'r'} 
\left( \pi \delta(\Omega_{p,p'q'r'}) -i {\cal P} \left( \frac{1}{\Omega_{p,p'q'r'}} \right) \right) 
d\pp d\pp' d\qq' d\rr'
$$
$$
+3 \sum_{s_{r} s_{p'} s_{q'} s_{r'}} \int_{\mathbb{R}^{4d}} \HH_{-\kk -\kk \rr -\rr} \HH_{-\kk \pp' \qq' \rr'} 
q_r^{s_r} q_{p'q'r'}^{-s_{k} s_{p'} s_{q'} s_{r'}} \delta_{k,p'q'r'} 
\left( \pi \delta(\Omega_{-k,p'q'r'}) -i {\cal P} \left( \frac{1}{\Omega_{-k,p'q'r'}} \right) \right) 
d\rr d\pp' d\qq' d\rr' .
$$
At this level, symmetries are very useful. First, we can show that the contributions of the products between second- and third-order cumulants cancel out exactly. Then, the remaining terms can be simplified using the symmetries of $\HH_{\kk \pp \qq \rr}$ and by playing with the dummy variables of integration. Finally, we find
\ba
{\partial q_k^{s_k} \over \partial T_{4}} &=& 
12 \pi \sum_{s_{p} s_{q} s_r} \int_{\mathbb{R}^{3d}} \HH_{\kk \pp \qq \rr} \HH_{-\kk -\pp -\qq -\rr} q_p^{s_p} q_q^{s_q} q_r^{s_r}
\delta_{k,pqr} \delta(\Omega_{k,pqr}) d\pp d\qq d\rr  \nonumber \\
&&+36 \pi \sum_{s_{p} s_{q} s_{r}} \int_{\mathbb{R}^{3d}} \HH_{\kk \pp \qq \rr} \HH_{\pp \kk -\qq -\rr} q_k^{s_k} q_q^{s_q} q_r^{s_r}  \delta_{k,pqr} 
\delta(\Omega_{k,pqr})  d\pp d\qq d\rr \nonumber \, .
\ea
A last simplification can be made by splitting the second term on the right hand side in three symmetrical contributions. We then obtain the kinetic equation for four-wave interactions (with $T_4 = \epsilon^4 t$)
\ba
{\partial q_k^{s_k} \over \partial t} &=& 
12 \epsilon^4 \pi \sum_{s_{p} s_{q} s_r} \int_{\mathbb{R}^{3d}} \HH_{\kk \pp \qq \rr} 
\left( - \frac{\HH_{\kk \pp \qq \rr}}{q_k^{s_k}} + \frac{\HH_{\pp \kk -\qq -\rr}}{q_p^{s_p}} + \frac{\HH_{\pp \kk -\pp -\rr}}{q_q^{s_q}} 
+ \frac{\HH_{\rr \kk -\pp -\qq}}{q_r^{s_r}} \right) q_k^{s_k} q_p^{s_p} q_q^{s_q} q_r^{s_r} \nonumber \\
&&\delta_{k,pqr} \delta(\Omega_{k,pqr}) d\pp d\qq d\rr \, .
\ea
This equation describes wave turbulence at the level of four-wave interactions. It is a much slower process than for three-wave interactions since it implies a time scale $T_4$ instead of $T_2$. As for three-wave interactions, it is the resonance mechanism which is at the origin of the redistribution of energy (or wave action).  

\end{appendix}
%Cited papers should include a DOI link. 

\bibliographystyle{SciPost_bibstyle}
\bibliography{WT-Biblio}

\begin{thebibliography}{10}
\providecommand{\url}[1]{\texttt{#1}}
\providecommand{\urlprefix}{URL }
\expandafter\ifx\csname urlstyle\endcsname\relax
  \providecommand{\doi}[1]{doi:\discretionary{}{}{}#1}\else
  \providecommand{\doi}{doi:\discretionary{}{}{}\begingroup
  \urlstyle{rm}\Url}\fi
\providecommand{\eprint}[2][]{\url{#2}}

\bibitem{GaltierCUP2023}
S.~{Galtier},
\newblock \emph{{Physics of Wave Turbulence}},
\newblock Cambridge University Press (2023).

\bibitem{Benney1966}
D.~{Benney} and P.~{Saffman},
\newblock \emph{{Nonlinear Interactions of Random Waves in a Dispersive
  Medium}},
\newblock Proc. R. Soc. Lond. A \textbf{289}(1418), 301 (1966),
\newblock \doi{10.1098/rspa.1966.0013}.

\bibitem{Phillips1981}
O.~{Phillips},
\newblock \emph{{Wave interactions - The evolution of an idea}},
\newblock J. Fluid Mech. \textbf{106}, 215 (1981),
\newblock \doi{10.1017/S0022112081001572}.

\bibitem{Stokes1847}
G.~{Stokes},
\newblock \emph{{On the theory of oscillatory waves}},
\newblock Trans. Camb. Philos. Soc. \textbf{8}, 441 (1847).

\bibitem{Peierls1929}
R.~{Peierls},
\newblock \emph{{Zur kinetischen Theorie der W{\"a}rmeleitung in Kristallen}},
\newblock Annalen der Physik \textbf{395}(8), 1055 (1929),
\newblock \doi{10.1002/andp.19293950803}.

\bibitem{Nordheim1928}
L.~{Nordheim},
\newblock \emph{{On the Kinetic Method in the New Statistics and Its
  Application in the Electron Theory of Conductivity}},
\newblock Proc. Royal Soc. London Series A \textbf{119}(783), 689 (1928),
\newblock \doi{10.1098/rspa.1928.0126}.

\bibitem{Phillips1960}
O.~{Phillips},
\newblock \emph{{On the dynamics of unsteady gravity waves of finite amplitude.
  Part 1. The elementary interactions}},
\newblock J. Fluid Mech. \textbf{9}, 193 (1960),
\newblock \doi{10.1017/S0022112060001043}.

\bibitem{Longuet1962}
M.~{Longuet-Higgins},
\newblock \emph{{Resonant interactions between two trains of gravity waves}},
\newblock J. Fluid Mech. \textbf{12}, 321 (1962),
\newblock \doi{10.1017/S0022112062000233}.

\bibitem{Hasselmann1962}
K.~{Hasselmann},
\newblock \emph{{On the non-linear energy transfer in a gravity-wave spectrum.
  Part 1. General theory}},
\newblock J. Fluid Mech. \textbf{12}, 481 (1962),
\newblock \doi{10.1017/S0022112062000373}.

\bibitem{Phillips1958}
O.~{Phillips},
\newblock \emph{{The equilibrium range in the spectrum of wind-generated
  waves}},
\newblock J. Fluid Mech. \textbf{4}, 426 (1958),
\newblock \doi{10.1017/S0022112058000550}.

\bibitem{Longuet1966}
M.~{Longuet-Higgins} and N.~{Smith},
\newblock \emph{{An experiment on third-order resonant wave interactions}},
\newblock J. Fluid Mech. \textbf{25}, 417 (1966),
\newblock \doi{10.1017/S0022112066000168}.

\bibitem{McGoldrick1966}
L.~{McGoldrick}, O.~{Phillips}, N.~{Huang} and T.~{Hodgson},
\newblock \emph{{Measurements of third-order resonant wave interactions}},
\newblock J. Fluid Mech. \textbf{25}, 437 (1966),
\newblock \doi{10.1017/S002211206600017X}.

\bibitem{Drummond1962}
W.~{Drummond} and D.~{Pines},
\newblock \emph{Non-linear stability of plasma oscillations},
\newblock Nuclear Fusion Supp. \textbf{3}, 1049 (1962).

\bibitem{2020Akylas}
T.~{Akylas},
\newblock \emph{{David J. Benney: Nonlinear Wave and Instability Processes in
  Fluid Flows}},
\newblock Ann. Rev. Fluid Mech. \textbf{52}(1), 010518 (2020),
\newblock \doi{10.1146/annurev-fluid-010518-040240}.

\bibitem{Benney1962}
D.~{Benney},
\newblock \emph{{Non-linear gravity wave interactions}},
\newblock J. Fluid Mech. \textbf{14}, 577 (1962),
\newblock \doi{10.1017/S0022112062001469}.

\bibitem{Poincare1893}
H.~{Poincar{\'e}},
\newblock \emph{{Les m{\'e}thodes nouvelles de la m{\'e}canique c{\'e}leste,
  vol. II}},
\newblock Dover publisher, New-York (1893).

\bibitem{Sturrock1957}
P.~{Sturrock},
\newblock \emph{{Non-Linear Effects in Electron Plasmas}},
\newblock Proc. R. Soc. Lond. A \textbf{242}(1230), 277 (1957),
\newblock \doi{10.1098/rspa.1957.0176}.

\bibitem{Nayfeh2004}
A.~{Nayfeh},
\newblock \emph{Perturbation Methods},
\newblock Wiley-VCH Verlag GmbH \& Co. KGaA, Weinheim (2004).

\bibitem{Benney1967}
D.~{Benney},
\newblock \emph{Asymptotic behavior of nonlinear dispersive waves},
\newblock J. Math. Phys. \textbf{46}(2), 115 (1967),
\newblock \doi{10.1002/sapm1967461115}.

\bibitem{Benney1967b}
D.~{Benney} and A.~{Newell},
\newblock \emph{Sequential time closures for interacting random waves},
\newblock J. Math. Phys. \textbf{46}(4), 363 (1967),
\newblock \doi{10.1002/sapm1967461363}.

\bibitem{Benney1969}
D.~{Benney} and A.~{Newell},
\newblock \emph{Random wave closures},
\newblock Stud. App. Maths. \textbf{48}(1), 29 (1969),
\newblock \doi{10.1002/sapm196948129}.

\bibitem{Sagdeev1966}
R.~{Sagdeev} and A.~{Galeev},
\newblock \emph{{Lectures on the non-linear theory of plasma}},
\newblock International Center for Theoretical Physics, Trieste (1966).

\bibitem{Vedenov67}
A.~{Vedenov},
\newblock \emph{{Theory of a Weakly Turbulent Plasma}},
\newblock Rev. Plasma Physics \textbf{3}, 229 (1967).

\bibitem{Kadomtsev1963}
B.~{Kadomtsev} and V.~{Petviashvili},
\newblock \emph{Weakly turbulent plasma in a magnetic field},
\newblock JETP \textbf{16}, 1578 (1963).

\bibitem{Zakharov1965}
V.~{Zakharov},
\newblock \emph{{Weak turbulence in media with a decay spectrum}},
\newblock J. Appl. Mech. Tech. Phys. \textbf{6}, 22 (1965),
\newblock \doi{10.1007/BF01565814}.

\bibitem{Zakharov1966}
V.~{Zakharov} and N.~{Filonenko},
\newblock \emph{{The energy spectrum for stochastic oscillations of a fluid
  surface}},
\newblock Doclady Akad. Nauk. SSSR \textbf{170}, 1292 (1966).

\bibitem{Zakharov1967}
V.~{Zakharov} and N.~{Filonenko},
\newblock \emph{{Weak turbulence of capillary waves}},
\newblock J. Applied Mech. Tech. Phys. \textbf{8}(5), 37 (1967),
\newblock \doi{10.1007/BF00915178}.

\bibitem{Zakharov1967b}
V.~{Zakharov},
\newblock \emph{{Weak-turbulence Spectrum in a Plasma Without a Magnetic
  Field}},
\newblock JETP \textbf{24}, 455 (1967).

\bibitem{Kaner1970}
{\'E}.~{Kaner} and V.~{Yakovenko},
\newblock \emph{{Weak Turbulence Spectrum and Second Sound in a Plasma}},
\newblock JETP \textbf{31}, 316 (1970).

\bibitem{Clark2014}
P.~{Clark di Leoni}, P.~{Cobelli} and P.~{Mininni},
\newblock \emph{{Wave turbulence in shallow water models}},
\newblock Phys. Rev. E \textbf{89}(6), 063025 (2014),
\newblock \doi{10.1103/PhysRevE.89.063025}.

\bibitem{BenneyNewell1967}
D.~J. {Benney} and A.~C. {Newell},
\newblock \emph{{Statistical Properties of the Sea}},
\newblock Phys. Fluids \textbf{10}(9), S281 (1967),
\newblock \doi{10.1063/1.1762469}.

\bibitem{Deike2011}
L.~{Deike}, C.~{Laroche} and E.~{Falcon},
\newblock \emph{{Experimental study of the inverse cascade in gravity wave
  turbulence}},
\newblock EPL (Europhys. Lett.) \textbf{96}(3), 34004 (2011),
\newblock \doi{10.1209/0295-5075/96/34004}.

\bibitem{Zhang2022}
Z.~{Zhang} and Y.~{Pan},
\newblock \emph{{Numerical investigation of turbulence of surface gravity
  waves}},
\newblock J. Fluid Mech. \textbf{933}, A58 (2022),
\newblock \doi{10.1017/jfm.2021.1114}.

\bibitem{Hwang2000}
P.~{Hwang}, D.~{Wang}, E.~{Walsh}, W.~{Krabill} and R.~{Swift},
\newblock \emph{{Airborne Measurements of the Wavenumber Spectra of Ocean
  Surface Waves. Part I: Spectral Slope and Dimensionless Spectral
  Coefficient*}},
\newblock J. Phys. Ocean. \textbf{30}(11), 2753 (2000),
\newblock \doi{10.1175/1520-0485(2001)031<2753:AMOTWS>2.0.CO;2}.

\bibitem{Lenain2017}
L.~{Lenain} and W.~{Melville},
\newblock \emph{{Measurements of the directional spectrum across the
  equilibrium saturation ranges of wind-generated surface waves}},
\newblock J. Phys. Ocean. \textbf{47}, 2123 (2017),
\newblock \doi{10.1175/JPO-D-17-0017.1}.

\bibitem{Falcon2022}
E.~{Falcon} and N.~{Mordant},
\newblock \emph{Experiments in surface gravity–capillary wave turbulence},
\newblock Ann. Rev. Fluid Mech. \textbf{54}(1), 1 (2022),
\newblock \doi{10.1146/annurev-fluid-021021-102043}.

\bibitem{MacKinnon2017}
J.~{MacKinnon} and {collaborators},
\newblock \emph{{Climate Process Team on Internal Wave-Driven Ocean Mixing}},
\newblock Bull. Am. Meteo. Soc. \textbf{98}(11), 2429 (2017),
\newblock \doi{10.1175/BAMS-D-16-0030.1}.

\bibitem{Caillol2000}
P.~{Caillol} and V.~{Zeitlin},
\newblock \emph{Kinetic equations and stationary energy spectra of weakly
  nonlinear internal gravity waves},
\newblock Dyn. Atmos. Oceans \textbf{32}(2), 81 (2000),
\newblock \doi{10.1016/S0377-0265(99)00043-3}.

\bibitem{Savaro2020}
C.~{Savaro}, A.~{Campagne}, M.~{Linares}, P.~{Augier}, J.~{Sommeria},
  T.~{Valran}, S.~{Viboud} and N.~{Mordant},
\newblock \emph{{Generation of weakly nonlinear turbulence of internal gravity
  waves in the Coriolis facility}},
\newblock Phys. Rev. Fluids \textbf{5}(7), 073801 (2020),
\newblock \doi{10.1103/PhysRevFluids.5.073801}.

\bibitem{Galtier2003}
S.~{Galtier},
\newblock \emph{{Weak inertial-wave turbulence theory}},
\newblock Phys. Rev. E \textbf{68}(1), 015301 (2003),
\newblock \doi{10.1103/PhysRevE.68.015301}.

\bibitem{Monsalve2020}
E.~{Monsalve}, M.~{Brunet}, B.~{Gallet} and P.-P. {Cortet},
\newblock \emph{Quantitative experimental observation of weak inertial-wave
  turbulence},
\newblock Phys. Rev. Lett. \textbf{125}, 254502 (2020),
\newblock \doi{10.1103/PhysRevLett.125.254502}.

\bibitem{LonguetH1967}
M.~{Longuet-Higgins} and A.~{Gill},
\newblock \emph{{Resonant Interactions between Planetary Waves}},
\newblock Proc. Royal Soc. Lond. A \textbf{299}(1456), 120 (1967),
\newblock \doi{10.1098/rspa.1967.0126}.

\bibitem{Balk1990}
A.~{Balk}, V.~{Zakharov} and S.~{Nazarenko},
\newblock \emph{{Nonlocal turbulence of drift waves}},
\newblock JETP \textbf{98}, 446 (1990).

\bibitem{Balk1990b}
A.~{Balk}, S.~{Nazarenko} and V.~{Zakharov},
\newblock \emph{{On the nonlocal turbulence of drift type waves}},
\newblock Phys. Lett. A \textbf{146}(4), 217 (1990),
\newblock \doi{10.1016/0375-9601(90)90168-N}.

\bibitem{Galtier2000}
S.~{Galtier}, S.~{Nazarenko}, A.~{Newell} and A.~{Pouquet},
\newblock \emph{{A weak turbulence theory for incompressible
  magnetohydrodynamics}},
\newblock J. Plasma Phys. \textbf{63}, 447 (2000),
\newblock \doi{10.1017/S0022377899008284}.

\bibitem{Finlay2008}
C.~{Finlay},
\newblock \emph{Waves in the presence of magnetic fields, rotation and
  convection},
\newblock In P.~Cardin and E.~s.~p. L.F. Cugliandolo~eds, eds., \emph{Dynamos},
  vol.~88 of \emph{Les Houches 2007}, pp. 403--450 (2008).

\bibitem{Galtier2014}
S.~{Galtier},
\newblock \emph{{Weak turbulence theory for rotating magnetohydrodynamics and
  planetary flows}},
\newblock J. Fluid Mech. \textbf{757}, 114 (2014),
\newblock \doi{10.1017/jfm.2014.490}.

\bibitem{Menu2019}
M.~{Menu}, S.~{Galtier} and L.~{Petitdemange},
\newblock \emph{{Inverse cascade of hybrid helicity in $B_{0}-\Omega$ MHD
  turbulence}},
\newblock Phys. Rev. Fluids \textbf{4}(7), 073701 (2019),
\newblock \doi{10.1103/PhysRevFluids.4.073701}.

\bibitem{Zakharov1970}
V.~{Zakharov} and R.~{Sagdeev},
\newblock \emph{{Spectrum of Acoustic Turbulence}},
\newblock Sov. Phys. Dok. \textbf{15}, 439 (1970).

\bibitem{Newell1971}
A.~{Newell} and P.~{Aucoin},
\newblock \emph{{Semi-dispersive wave systems}},
\newblock J. Fluid Mech. \textbf{49}, 593 (1971),
\newblock \doi{10.1017/S0022112071002271}.

\bibitem{Lvov1997}
V.~{L'vov}, Y.~{L'vov}, A.~{Newell} and V.~{Zakharov},
\newblock \emph{{Statistical description of acoustic turbulence}},
\newblock Phys. Rev. E \textbf{56}(1), 390 (1997),
\newblock \doi{10.1103/PhysRevE.56.390}.

\bibitem{Galtier2023}
S.~{Galtier},
\newblock \emph{{Fast magneto-acoustic wave turbulence and the
  Iroshnikov-Krachnan spectrum}},
\newblock J. Fluid Mech. \textbf{89}(2), 905890205 (2023),
\newblock \doi{10.1017/S0022377823000259}.

\bibitem{Steinberg2021}
V.~{Steinberg},
\newblock \emph{{Elastic Turbulence: An Experimental View on Inertialess Random
  Flow}},
\newblock Ann. Rev. Fluid Mech. \textbf{53}(1), 010719 (2021),
\newblock \doi{10.1146/annurev-fluid-010719-060129}.

\bibitem{During2006}
G.~{D{\"u}ring}, C.~{Josserand} and S.~{Rica},
\newblock \emph{{Weak Turbulence for a Vibrating Plate: Can One Hear a
  Kolmogorov Spectrum?}},
\newblock Phys. Rev. Lett. \textbf{97}(2), 025503 (2006),
\newblock \doi{10.1103/PhysRevLett.97.025503}.

\bibitem{During2015}
G.~{D{\"u}ring}, C.~{Josserand} and S.~{Rica},
\newblock \emph{{Self-similar formation of an inverse cascade in vibrating
  elastic plates}},
\newblock Phys. Rev. E \textbf{91}(5), 052916 (2015),
\newblock \doi{10.1103/PhysRevE.91.052916}.

\bibitem{Boudaoud2008}
A.~{Boudaoud}, O.~{Cadot}, B.~{Odille} and C.~{Touz{\'e}},
\newblock \emph{{Observation of Wave Turbulence in Vibrating Plates}},
\newblock Phys. Rev. Lett. \textbf{100}(23), 234504 (2008),
\newblock \doi{10.1103/PhysRevLett.100.234504}.

\bibitem{Mordant2008}
N.~{Mordant},
\newblock \emph{{Are There Waves in Elastic Wave Turbulence?}},
\newblock Phys. Rev. Lett. \textbf{100}(23), 234505 (2008),
\newblock \doi{10.1103/PhysRevLett.100.234505}.

\bibitem{Cobelli2009}
P.~{Cobelli}, P.~{Petitjeans}, A.~{Maurel}, V.~{Pagneux} and N.~{Mordant},
\newblock \emph{{Space-Time Resolved Wave Turbulence in a Vibrating Plate}},
\newblock Phys. Rev. Lett. \textbf{103}(20), 204301 (2009),
\newblock \doi{10.1103/PhysRevLett.103.204301}.

\bibitem{Mordant2010}
N.~{Mordant},
\newblock \emph{{Fourier analysis of wave turbulence in a thin elastic plate}},
\newblock Europ. Phys. J. B \textbf{76}(4), 537 (2010),
\newblock \doi{10.1140/epjb/e2010-00197-y}.

\bibitem{Sulem1999}
C.~{Sulem} and P.-L. {Sulem},
\newblock \emph{Nonlinear Schr\"odinger Equation: Self-Focusing and Wave
  Collapse}, vol. 139,
\newblock Springer-Verlag, New-York (1999).

\bibitem{Dyachenko1992}
S.~{Dyachenko}, A.~{Newell}, A.~{Pushkarev} and V.~{Zakharov},
\newblock \emph{{Optical turbulence: weak turbulence, condensates and
  collapsing filaments in the nonlinear Schrodinger equation}},
\newblock Physica D \textbf{57}, 96 (1992),
\newblock \doi{10.1016/0167-2789(92)90090-A}.

\bibitem{Laurie2012}
J.~{Laurie}, U.~{Bortolozzo}, S.~{Nazarenko} and S.~{Residori},
\newblock \emph{{One-dimensional optical wave turbulence: Experiment and
  theory}},
\newblock Phys. Rep. \textbf{514}(4), 121 (2012),
\newblock \doi{10.1016/j.physrep.2012.01.004}.

\bibitem{Mitchell1996}
M.~{Mitchell}, Z.~{Chen}, M.-F. {Shih} and M.~{Segev},
\newblock \emph{{Self-Trapping of Partially Spatially Incoherent Light}},
\newblock Phys. Rev. Lett. \textbf{77}(3), 490 (1996),
\newblock \doi{10.1103/PhysRevLett.77.490}.

\bibitem{Picozzi2014}
A.~{Picozzi}, J.~{Garnier}, T.~{Hansson}, P.~{Suret}, S.~{Randoux}, G.~{Millot}
  and D.~{Christodoulides},
\newblock \emph{{Optical wave turbulence: Towards a unified nonequilibrium
  thermodynamic formulation of statistical nonlinear optics}},
\newblock Phys. Rep. \textbf{542}(1), 1 (2014),
\newblock \doi{10.1016/j.physrep.2014.03.002}.

\bibitem{Henn2009}
E.~{Henn}, J.~{Seman}, G.~{Roati}, K.~{Magalhaes} and V.~{Bagnato},
\newblock \emph{{Emergence of Turbulence in an Oscillating Bose-Einstein
  Condensate}},
\newblock Phys. Rev. Lett. \textbf{103}(4), 045301 (2009),
\newblock \doi{10.1103/PhysRevLett.103.045301}.

\bibitem{Nazarenko2006BEC}
S.~{Nazarenko} and M.~{Onorato},
\newblock \emph{{Wave turbulence and vortices in Bose Einstein condensation}},
\newblock Physica D \textbf{219}(1), 1 (2006),
\newblock \doi{10.1016/j.physd.2006.05.007}.

\bibitem{Nazarenko2007}
S.~{Nazarenko} and M.~{Onorato},
\newblock \emph{{Freely decaying Turbulence and Bose Einstein Condensation in
  Gross Pitaevski Model}},
\newblock J. Low Temp. Phys. \textbf{146}(1-2), 31 (2007),
\newblock \doi{10.1007/s10909-006-9271-z}.

\bibitem{Proment2009}
D.~{Proment}, S.~{Nazarenko} and M.~{Onorato},
\newblock \emph{{Quantum turbulence cascades in the Gross-Pitaevskii model}},
\newblock Phys. Rev. A \textbf{80}(5), 051603 (2009),
\newblock \doi{10.1103/PhysRevA.80.051603}.

\bibitem{Proment2012}
D.~{Proment}, S.~{Nazarenko} and M.~{Onorato},
\newblock \emph{{Sustained turbulence in the three-dimensional Gross-Pitaevskii
  model}},
\newblock Physica D \textbf{241}(3), 304 (2012),
\newblock \doi{10.1016/j.physd.2011.06.007}.

\bibitem{Vinen2000}
W.~{Vinen},
\newblock \emph{{Classical character of turbulence in a quantum liquid}},
\newblock Phys. Rev. B \textbf{61}(2), 1410 (2000),
\newblock \doi{10.1103/PhysRevB.61.1410}.

\bibitem{Kivotides2001}
D.~{Kivotides}, J.~{Vassilicos}, D.~{Samuels} and C.~{Barenghi},
\newblock \emph{{Kelvin Waves Cascade in Superfluid Turbulence}},
\newblock Phys. Rev. Lett. \textbf{86}(14), 3080 (2001),
\newblock \doi{10.1103/PhysRevLett.86.3080}.

\bibitem{Kozik2004}
E.~{Kozik} and B.~{Svistunov},
\newblock \emph{{Kelvin-Wave Cascade and Decay of Superfluid Turbulence}},
\newblock Phys. Rev. Lett. \textbf{92}(3), 035301 (2004),
\newblock \doi{10.1103/PhysRevLett.92.035301}.

\bibitem{Nazarenko2006}
S.~{Nazarenko},
\newblock \emph{{Differential approximation for Kelvin Wave Turbulence}},
\newblock JETP Lett. \textbf{83}(5), 198 (2006),
\newblock \doi{10.1134/S0021364006050031}.

\bibitem{Galtier2020gw}
S.~{Galtier}, J.~{Laurie} and S.~{Nazarenko},
\newblock \emph{{A plausible model of inflation driven by strong gravitational
  wave turbulence}},
\newblock Universe \textbf{6}(7), 98 (2020),
\newblock \doi{10.3390/universe6070098}.

\bibitem{Galtier2017}
S.~{Galtier} and S.~{Nazarenko},
\newblock \emph{{Turbulence of Weak Gravitational Waves in the Early
  Universe}},
\newblock Phys. Rev. Lett. \textbf{119}(22), 221101 (2017),
\newblock \doi{10.1103/PhysRevLett.119.221101}.

\bibitem{Galtier2021}
S.~{Galtier} and S.~{Nazarenko},
\newblock \emph{{Direct Evidence of a Dual Cascade in Gravitational Wave
  Turbulence}},
\newblock Phys. Rev. Lett. \textbf{127}(13), 131101 (2021),
\newblock \doi{10.1103/PhysRevLett.127.131101}.

\bibitem{Hassaini2019}
R.~{Hassaini}, N.~{Mordant}, B.~{Miquel}, G.~{Krstulovic} and G.~{D{\"u}ring},
\newblock \emph{{Elastic weak turbulence: From the vibrating plate to the
  drum}},
\newblock Phys. Rev. E \textbf{99}(3), 033002 (2019),
\newblock \doi{10.1103/PhysRevE.99.033002}.

\bibitem{Deng2021}
Y.~{Deng} and H.~{Zaher},
\newblock \emph{On the derivation of the wave kinetic equation for nls},
\newblock Forum Math. Pi \textbf{9}, 1 (2021),
\newblock \doi{doi:10.1017/fmp.2021.6}.

\bibitem{Connaughton2001}
C.~{Connaughton}, S.~{Nazarenko} and A.~{Pushkarev},
\newblock \emph{Discreteness and quasiresonances in weak turbulence of
  capillary waves},
\newblock Phys. Rev. E \textbf{63}, 046306 (2001),
\newblock \doi{10.1103/PhysRevE.63.046306}.

\bibitem{Bouroudia2008}
L.~{Bourouiba},
\newblock \emph{Discreteness and resolution effects in rapidly rotating
  turbulence},
\newblock Phys. Rev. E \textbf{78}, 056309 (2008),
\newblock \doi{10.1103/PhysRevE.78.056309}.

\bibitem{Galtier2020}
S.~{Galtier},
\newblock \emph{{Wave turbulence: the case of capillary waves}},
\newblock Geophys. Astro. Fluid Dyn. \textbf{115}(3), 234 (2021),
\newblock \doi{10.1080/03091929.2020.1715966}.

\bibitem{David2023}
V.~{David} and S.~{Galtier},
\newblock \emph{{Locality of triadic interaction and Kolmogorov constant in
  inertial wave turbulence:}},
\newblock J. Fluid Mech. Rapids \textbf{955}, R2 (2023).

\bibitem{LeReun2020}
T.~{Le Reun}, B.~{Favier} and M.~{Le Bars},
\newblock \emph{{Evidence of the Zakharov-Kolmogorov spectrum in numerical
  simulations of inertial wave turbulence}},
\newblock EPL (Europhysics Letters) \textbf{132}(6), 64002 (2020).

\bibitem{Yokoyama2021}
N.~{Yokoyama} and M.~{Takaoka},
\newblock \emph{{Energy-flux vector in anisotropic turbulence: application to
  rotating turbulence}},
\newblock J. Fluid Mech. \textbf{908}, A17 (2021).

\bibitem{Deng2023}
Y.~{Deng} and Z.~{Hani},
\newblock \emph{Full derivation of the wave kinetic equation},
\newblock Inventiones Mathematicae pp. 1--182 (2023),
\newblock \doi{10.1007/s00222-023-01189-2}.

\bibitem{Nazarenko11}
S.~{Nazarenko},
\newblock \emph{Wave Turbulence}, vol. 825 of \emph{Lecture Notes in Physics},
\newblock Berlin Springer Verlag (2011).

\end{thebibliography}
\nolinenumbers
\end{document}